\documentclass[10pt,letterpaper]{article}
\usepackage[margin=1in,nohead,centering]{geometry}
\usepackage{appendix}
\usepackage{framed}
\usepackage{amssymb}
\usepackage{amsmath}
\usepackage{url}
\usepackage{xspace}
\usepackage{graphics,graphicx}

\newcommand{\mathbbp}{{\bf p}}
\newcommand{\mathbbP}{{\bf P}}
\newcommand{\mathbbQ}{{\bf Q}}

\newtheorem{theorem}{Theorem}
\newtheorem{algorithm}{Algorithm}
\newtheorem{remark}{Remark}

\newtheorem{example}{Example}
\newtheorem{corollary}{Corollary}
\newtheorem{definition}{Definition}

\newcommand{\op}{\,\mbox{\bf\texttt{(}}\,}
\newcommand{\cp}{\,\mbox{\bf\texttt{)}}\,}

\newcommand{\aseq}{{\bf a}\xspace}

\chardef\other=12
\def\mdeactivate{%
\catcode`\&=\other   \catcode`\#=\other
\catcode`\%=\other   \catcode`\~=\other
}
\def\mmakeactive#1{\catcode`#1=\active\ignorespaces}

{
\mmakeactive\
\gdef\obeywhitespace{%
  \mmakeactive\^^M %
  \let^^M=\NewLine %
  \aftergroup\removebox %
  \obeyspaces %
}}

\def\NewLine{\par\indent}
\def\removebox{\setbox0=\lastbox}

\def\mverbatim{\par\begingroup\parindent=0em\tt\mdeactivate\obeywhitespace
\catcode`\|=0  %
}

\def\|{|}

\title{RNA folding kinetics using Monte Carlo and Gillespie 
algorithms\thanks{Research supported by
National Science Foundation grant DBI-1262439.}}
\author{Peter Clote \and Amir H. Bayegan}
\date{Department of Biology, Boston  College,
Chestnut Hill, MA 02467, USA.  {\tt clote@bc.edu}.}

\begin{document}
\maketitle

\begin{abstract}
RNA secondary structure folding kinetics is known to be
important for the biological function of certain processes, such as
the hok/sok system in {\em E. coli}.  Although linear algebra provides an exact
computational solution of secondary structure folding kinetics with respect to
the Turner energy model for tiny ($\approx 20$ nt)
RNA sequences, the folding kinetics for larger sequences can
only be approximated by binning structures into macrostates in a 
coarse-grained model, or by repeatedly simulating secondary structure 
folding with either the Monte Carlo algorithm or the Gillespie algorithm.

Here we investigate the relation between the Monte Carlo algorithm and
the Gillespie algorithm.  We prove that asymptotically, the
{\em expected time} for a $K$-step trajectory of the Monte Carlo algorithm
is equal to $\langle N \rangle$ times that of the Gillespie algorithm,
where $\langle N \rangle$ denotes the Boltzmann expected {\em network degree}.
If the network is regular (i.e. every node has the same degree),
then the {\em mean first passage time} (MFPT) computed by the Monte Carlo
algorithm is equal to MFPT computed by the Gillespie algorithm multiplied
by $\langle N \rangle$; however, this is not true for non-regular networks.
In particular, RNA secondary structure folding kinetics, as computed
by the Monte Carlo algorithm, is not equal to the folding kinetics, as 
computed by the Gillespie algorithm, although the mean first passage times
are roughly correlated.

Simulation software for RNA secondary structure folding according to 
the Monte Carlo and Gillespie algorithms is publicly available, as is our
software to compute the expected degree of the network of 
secondary structures of a given RNA sequence -- see
\url{http://bioinformatics.bc.edu/clote/RNAexpNumNbors}.
\end{abstract}

\section{Introduction}
\label{section:introduction}

Ever since Anfinsen's pioneering experiment involving the denaturation and
renaturation of bovine ribonuclease \cite{anfinsen}, 
it has long been held that the native state of a protein is that conformation
which has minimum free energy (MFE) among all possible conformations.
For that reason, both the
Monte Carlo algorithm \cite{metropolis:MonteCarlo} and the
Gillespie algorithm \cite{gillespieStochasticSimulation1} have
been used in biopolymer folding studies to estimate 
{\em kinetic folding time}, even when {\em detailed balance} does not 
hold.\footnote{Definitions of Monte Carlo
algorithm, Gillespie algorithm, detailed balance, mean first passage time,
master equation, etc. are given later in the paper.} 
Since a number of different measures
of folding time have been described for RNA secondary structure formation
\cite{Schmitz.jmb96,flamm,Wolfinger:04a,Senter.po12}, the goal of the 
current paper is to clarify the relations between these measures, to
point out the possibly low correlation between mean first passage time 
(MFPT) when computed by the Monte Carlo algorithm versus 
Gillespie algorithm, and to prove an asymptotic result that 
precisely relates Monte Carlo trajectory time with Gillespie trajectory time.
We hope that this clarification will be useful to those users who
perform computational experiments to estimate RNA folding kinetics.

\subsubsection*{Protein folding kinetics using Monte Carlo and/or Gillespie}

For protein to be biologically useful, its
amino acid sequence must satisfy two requirements: (1)
a {\em thermodynamic} requirement that the sequence have a unique, 
thermodynamically stable native structure, and (2) a {\em kinetic} 
requirement that the denatured polypeptide chain be capable of reaching 
the native state within reasonable time (milliseconds to seconds) under 
appropriate solution conditions. 
In \cite{Sali.n94,karplus:JMolecBiol94} 
a simple Monte Carlo folding experiment was designed for a 27 bead 
hetero-polymer on a cubic lattice using hydrophobic contact potentials.
Defining a sequence to fold rapidly provided its MFPT is sufficiently small,
the authors concluded that a necessary and sufficient condition
for a sequence to fold rapidly in their model is the existence of
a pronounced energy minimum with substantial gap between the energy of
the native state and that of the most stable misfolded structure.
A succinct summary of the theoretical work in
\cite{Sali.n94,karplus:JMolecBiol94} 
is that the thermodynamic requirement entails the kinetic requirement.
In \cite{Abkevich.pnas96,Abkevich.psb97} Monte Carlo hetero-polymer
folding experiments were performed to show how prebiotic proteins might
have evolved to fold rapidly in a primordial soup in order to survive
hydrolysis. Hetero-polymer sequences were {\em evolved} by applying 
a Metropolis criterion to the acceptance of pointwise mutations, depending
on whether the mutation increased MFPT. One of the conclusions from this
work was that the kinetic requirement entails the thermodynamic requirement.

More recently, the notion of {\em Markov state model} (MSM) has been introduced
to analyze protein folding kinetics by statistical analysis of a number of
molecular dynamics (MD) trajectories -- see for instance
\cite{Swope,Bowman.m09,Huang.psb10,Weber.jctc11,Bowman.mmb14,Harrigan.bj17}.
To construct a Markov state model, microstates are first formed by applying
a clustering algorithm to conformations taken at fixed time intervals 
(shapshots) from MD trajectories by using a measure of 
structural similarity such as RMSD. States of the Markov state model
are then defined to be kinetically reachable macrostates constructed from
microstates, where transition probabilities are defined by counting the
relative number of transitions between macrostates.  The kinetics of folding
are then analyzed by computing MFPT to reach the macrostate of the MSM 
which contains the native state.  Provided that the number of states 
is reasonably small, the MFPT can be computed analytically from the 
{\em fundamental matrix} as done in \cite{Weber.jctc11}. Analytic computations
of MFPT are much more efficient and accurate than repeated simulations
of the Monte Carlo algorithm -- however, both methods are theoretically
equivalent. In contrast, in some papers such as \cite{Zhang.pnas02},
instead of computing the MFPT of the Markov state model, transition rates
are computed from which the {\em master equation} is solved to determine
the {\em equilibrium time}, i.e. time necessary for
molecular {\em equilibrium}. Provided that the number of states is
reasonably small, it is possible to solve the master equation analytically
by spectral decomposition; otherwise, the Gillespie algorithm 
can be used to stochastically estimate the equilibrium time.
In \cite{Levy.ps13}, a formula (involving the integral of an appropriate time
correlation function) is given to estimate the
equilibrium time among unfolded states of a protein having
two state kinetics.

\subsubsection*{RNA folding kinetics using Monte Carlo and/or Gillespie}

In \cite{Schmitz.jmb96}, the {\em folding time} for an RNA sequence is 
defined to be the sum of reciprocals of the rate constants 
$k = k_0 \cdot \exp(-\Delta E^{\ddag}/RT)$ in each step of a folding 
trajectory, where $k_0$ is a rate calibration constant and $\Delta E^{\ddag}$ 
is the activation energy of adding or removing a base pair to the current
structure, using the Turner free energy model \cite{Turner.nar10}. 
The resultant folding time differs from MFPT by additionally accounting
for the transition rates between trajectory steps. A similar notion of
folding time is adopted in \cite{Geis.jmb08}, where additionally
basins of attraction are estimated by assigning a collection of
sampled secondary structures to a locally optimal structure by a greedy
procedure.

In \cite{Senter.po12}, mean first passage time was estimated by 
the average number of Monte Carlo steps, taken over 50 runs, to fold the 
empty secondary structure into the MFE structure.  The authors determined
that for the selenocysteine (SECIS-1) family RF00031 from
the Rfam database \cite{Nawrocki.nar15}, 
the Pearson correlation coefficient is $0.48436$ 
(p-value $1.8 \cdot 10^{-4}$) between the logarithm of MFPT for a given RNA
sequence and the standard deviation of the number of base pairs taken 
over the ensemble of all secondary structures of that sequence. 
After publication of \cite{Senter.po12}, computational experiments 
performed with the Gillespie algorithm in place of the Monte Carlo algorithm
by using {\tt Kinfold} \cite{flamm} suggested that there is no 
such correlation between log MFPT and standard deviation of the number of 
base pairs of secondary structures in the ensemble (data not shown). 
This observation raises the question of which method to use for RNA 
folding kinetics.

In \cite{flamm} the program {\tt Kinfold} is described, now part of
Vienna RNA Package \cite{Lorenz.amb11}, which implements
the Gillespie algorithm for RNA secondary structure formation for the
Turner energy model \cite{Turner.nar10}. In that paper,
{\em folding time} is defined to be the {\em first passage time} from
the empty structure to a given target structure, so that the MFPT is
expected folding time.  In \cite{Aviram.amb12}, an increase in
speed of {\tt Kinfold} is reported, by modifying the source code to 
incorporate memoization. In \cite{Dykeman.nar15}, the program
{\tt KFOLD} is described, which also implements the Gillespie algorithm,
uses the Turner energy model, and performs
the same elementary step base pair addition/removal transitions 
as does {\tt Kinfold}; however, {\tt KFOLD} achieves a
remarkable speed-up over {\tt Kinfold} by exploiting the fact that many of 
the base-pair addition/deletion moves and their corresponding
rates do not change between each step in the simulation. As in 
\cite{flamm}, folding time is defined in \cite{Dykeman.nar15} to be the
{\em first passage time} from the empty structure to a given target structure.

Instead of determining the mean first passage time for RNA secondary
structure formation by using the Monte Carlo algorithm or the 
Gillespie algorithm, an alternative approach is to compute 
the {\em equilibrium time} by solving the master equation or by
simulating the Gillespie algorithm until equilibrium has been achieved.
In \cite{Wolfinger:04a} the method {\tt Treekin} is described, where
macrostates are defined by {\em basins of attraction}, and
population occupancy curves are computed over time for each macrostate.
A similar approach is described in \cite{Senter.jmb15}, where the
authors use a different method of constructing macrostates.

\subsubsection*{Relation between MFPT for Markov chain versus process}

In this paper, we consider {\em mean first passage time} (MFPT) of
random walks in a finite, discrete-time {\em Markov chain} $\mathbb{M}_1$, 
with transition probabilities given by 
equation~(\ref{eqn:transitionProb1}) in the next section.
We also consider MFPT of trajectories in the finite, continuous-time
{\em Markov process} $\mathbb{M}_2$,
with transition rates given by equation~(\ref{eqn:transitionProb2}) in the
next section. Mean first passage
time for the Markov chain $\mathbb{M}_1$ is computed by repeated simulations of
the Metropolis Monte Carlo algorithm, while
that for the Markov process $\mathbb{M}_2$ is computed by repeated 
simulations of the Gillespie algorithm.
Although distinct, both methods to compute MFPT have been described 
in the literature -- see summary in the previous two sections.
There appears to be a tacit and sometimes explicit assumption
that the kinetics results are essentially equivalent 
\cite{amatoRecombRNA,Tang.jmb08}.

Indeed, mean first passage times for protein and/or RNA folding have been
computed by matrix inversion and/or Monte Carlo simulations over a Markov
chain \cite{Klemm08funnelsin,Tang.jmb08,Bowman.pnas10,Huang.psb10,Weber.jctc11,Levy.ps13}, while the master equation and/or Gillespie simulations have 
been used in \cite{flamm,Wolfinger:04a,Aviram.amb12,Dykeman.nar15,Xu.pnas16}. 
Of particular interest is the construction of Markov state models from
molecular dynamics folding trajectories for a protein or RNA molecule,
followed by either mean first passage time computed by matrix inversion
\cite{Weber.jctc11} versus equilibrium time computed by solving the
master equation \cite{Xu.pnas16}.
This leads to the question: what is the relation between
mean first passage time computed by the Monte Carlo algorithm versus
the Gillespie algorithm?

The answer to this question is trivial in the case that
the Markov chain is $N$-regular, i.e. that each state has degree $N$
($N$ neighbors). 
In this case, we show 
the MFPT obtained by the time-driven Monte Carlo algorithm is
equal to $N$ times that of the Gillespie algorithm.
In the context of RNA secondary structures of a
given sequence RNA, this
suggests that MFPT determined by the time-driven Monte Carlo algorithm 
might be approximately equal to $\langle N \rangle$ times the MFPT
determined by the Gillespie algorithm, where $\langle N \rangle$ 
denotes the Boltzmann expected node {\em degree}
(expected number of structural neighbors), formally defined by
\begin{eqnarray}
\label{eqn:networkDegree}
\langle N \rangle &=& \sum_{s} N(s) \cdot P(s) =
\sum_{s} N(s) \cdot \frac{\exp(-E(s)/RT)}{Z}
\end{eqnarray}
where the sum is taken over all secondary structures $s$ of the RNA
sequence, $N(s)$ is the number of neighbors of $s$, i.e. structures that
can be obtained from $s$ by the addition or removal of a single base pair,
$E(s)$ is the free energy of structure $s$ using the Turner energy model
\cite{Turner.nar10}, $R$ the universal gas constant, $T$ absolute 
temperature, and $Z = \sum_{s} \exp(-E(s)/RT)$ denotes the partition function. 
However, we show that this is not the case, 
and can only conclude that the Monte Carlo and Gillespie algorithms 
return loosely correlated mean first passage times for RNA secondary
structure folding. On the other hand,
we prove that asymptotically, the {\em expected time} for a $K$-transition
Monte Carlo trajectory is equal to  that for a $K$-transition
Gillespie trajectory multiplied by $\langle N \rangle$.

A rapidly emerging area of synthetic biology concerns the
computational design of synthetic RNA \cite{Choi.z14,Dotu.nar15}
by using inverse folding software
\cite{Taneda.aabc11,Zadeh.jcc11,GarciaMartin13,EsmailiTaheri.bb15}.
It seems clear that the next step in synthetic RNA design
will be to control the kinetics of RNA folding. Due to the differences
between MFPT computations with the Monte Carlo and Gillespie algorithms
shown in this paper, we suggest that the Gillespie algorithm and
related population occupancy curves determined by solution of the
{\em master equation}
constitute a more appropriate approach to macromolecular folding
kinetics \cite{Xu.pnas16}, rather than the Monte Carlo algorithm and
related computation of mean first passage time by matrix inversion
\cite{Bowman.pnas10}. In the context of synthetic RNA design, we advocate
the computation of MFPT using matrix inversion for a coarse-grained model
\cite{Senter.jcb15}, as an efficient, initial screen of
potential candidate, followed by the slower but more accurate coarse-grain
method {\tt Treekin} \cite{flamm,wolfingerStadler:kinetics}, followed by
repeated simulations using {\tt KFOLD} \cite{Dykeman.nar15}.

\subsubsection*{Plan of the paper}

In Section~\ref{section:definitions},
we provide some background on RNA secondary structure,
Markov chains and Markov processes. 
Section~\ref{section:HTandET} describes three versions of the Monte Carlo
algorithm (discrete-time time-driven, discrete-time event-driven and
continuous-time event-driven) and the Gillespie algorithm, and illustates
these algorithms with a simple example. Readers familiar with Markov chains,
Markov processes, Monte Carlo algorithm and the Gillespie algorithm should
read the pseudocode of Algorithms 1-4 (given in Figures 1-4) and otherwise
skip Sections~\ref{section:definitions},\ref{section:HTandET} and proceed
directly to Section~\ref{section:results}. That section
presents a proof of a new theorem that Monte Carlo trajectory 
time asymptotically equals Gillespie trajectory time multiplied 
by the expected degree. RNA secondary structure folding simulations using
the Monte Carlo algorithm and the Gillespie algorithm are given, which
provide a computational illustration of the theorem. Despite the close
asymptotic relation between Monte Carlo and Gillespie trajectory time,
Section~\ref{section:results} shows that there is no such relation
between Monte Carlo and Gillespie mean first passage time for RNA
secondary structure folding, but only a loose correlation.
Section~\ref{section:discussion} presents some discussion and
concluding remarks. The Appendix presents computations that show
that detailed balance does not hold for the Markov chain of secondary
structures of an RNA sequence.

\section{Background}
\label{section:definitions}

In this section, we describe some definitions concerning RNA secondary
structures, expected degree of a secondary structure network, 
Markov chains and Markov processes. We use the notation
$\mathbb{R}$ for the set of real numbers,
$\mathbb{R}^{\geq 0}$ for the set of non-negative real numbers,
and $\mathbb{N}$ for the set of natureal numbers $0,1,2,\ldots$.

\subsection{Background on RNA secondary structure}

A secondary structure for an RNA nucleotide sequence
$\aseq = a_1,\dots,a_n$ is a set $s$ of Watson-Crick or wobble
base pairs $(i,j)$, containing neither base triples nor pseudoknots.
More formally we have the following definition.
\begin{definition}
\label{def:secStr}
A secondary structure for a given RNA nucleotide sequence
$a_1,\dots,a_n$ is a set $s$ of base pairs $(i,j)$, where $1 \leq i < j \leq n$,
such that:
\begin{enumerate}
\item
if $(i,j)\in s$ then
$a_i,a_j$ form either a Watson-Crick (AU,UA,CG,GC) or
wobble (GU,UG) base pair,
\item
if $(i,j)\in s$ then $j-i>\theta=3$ (a steric constraint
requiring that there be at least $\theta=3$ unpaired bases between
any two positions that are paired),
\item
if $(i,j)\in s$ then for all $i' \ne i$ and $j' \ne j$,
$(i',j) \not\in s$ and $(i,j') \not\in s$ (nonexistence of base triples),
\item
if $(i,j)\in s$ and $(k,\ell)\in s$, then
it is not the case that $i<k<j<\ell$ (nonexistence of pseudoknots).
\end{enumerate}
\end{definition}

The Turner energy model \cite{Turner.nar10} is an additive energy
model, where enthalpies, entropies and free energies
for stacked base pairs, hairpins, bulges, internal loops, etc.
are derived by least-squares fitting of experimental data 
from UV-absorbance (so-called {\em optical melting}) experiments,
following the pioneering work of Tinoco \cite{tinoco:reviewInBook}.
In this model, there is no energy contribution for a base pair; in contrast,
stacked base pairs contribute negative, stabilizing free energy, while
loop regions contribute positive, destabilizing free energy due to entropic
loss.  For instance, from melting experiments at $37^{\circ}$ C,
Turner's rules assign stacking free energy of $–2.08 \pm 0.06$ and
enthalpy of $–10.48 \pm 1.24$ to the stacked pair
$\begin{array}{ll}
\mbox{$5'$-{\tt CU}-$3'$}\\
\mbox{$3'$-{\tt GA}-$5'$}\\
\end{array}$
The minimum free energy (MFE) secondary structure can be computed in time
that is cubic with respect to the length of an input RNA sequence; this is
done by {\em dynamic programming} following the Zuker algorithm 
\cite{zukerStiegler}, as implemented in publicly available software, such
as the Vienna RNA Package \cite{Lorenz.amb11}.

Secondary structures can be depicted in several equivalent manners.
For instance, the sequence and dot-bracket representation for 
a type III hammerhead ribozyme 
from Peach Latent Mosaic Viroid (PLMVd) AJ005312.1/282-335
is given by
\begin{quote}
\bf
\begin{verbatim}
GAUGAGUCUGUGCUAAGCACACUGAUGAGUCUAUGAAAUGAGACGAAACUCAUA
.((((((.(((((...))))).......((((........))))...)))))).
\end{verbatim}
\end{quote}
The left panel of Figure~\ref{fig:Vienna185andVienna207} displays the
equivalent usual presentation of the same structure, computed by {\tt RNAfold} 
from the Vienna RNA Package \cite{Lorenz.amb11} using the Turner 1999 
energy parameters.  This minimum free energy (MFE) structure agrees with the 
consensus structure from the Rfam database \cite{Gardner.nar11}.  The right 
panel of Figure~\ref{fig:Vienna185andVienna207} depicts the network of all
secondary structures for the 11-nt RNA sequence {\tt GCGCGCGCGCG}, containing
27 nodes and 42 edges, where nodes are secondary structures in dot-bracket
notation, and edges are indicated between each two structures $s,t$ in which
$t$ is obtained by adding or removing a single base pair to/from $s$.
The move set $MS_1$ [resp. $MS_2$] consists of adding or removing
[resp. adding, removing or shifting \cite{flamm,Clote.po15}] a base pair
(in this paper, we mostly concentrate on the $MS_1$ move set).
{\em Mean first passage time} (MFPT) for $\mathcal{M}_1$
[resp. $\mathcal{M}_2$] is the average first passage time from 
a designated initial state $x_0$ to a designated final state $x_{\infty}$,
which can be approximated by averaging over a number of runs of the
Monte Carlo algorithm [resp. Gillespie algorithm]. In the context of
RNA secondary structure folding kinetics, the initial structure 
$x_0$ is often taken to be the empty structure and the target structure
$x_{\infty}$ is often taken to be the MFE structure -- see
Figure~\ref{fig:Vienna185andVienna207}) where the empty structure is
indicated by a green circle, and the MFE structure by an orange circle.
The contribution of
this paper is to show that asymptotic trajectory times for Monte Carlo and
Gillespie algorithms are related, but that mean first passage times appear
to be only loosely correlated.

\subsection{Background on Markov chains and processes}
Suppose that $X=\{1,\ldots,n\}$ 
is a finite set of states, and for each state $x$ there
is an associated energy $E(x)$. Suppose that $G$ is a graph with vertex
set $X$ and an arbitrary, but fixed edge set. $G$ could be a complete 
graph with edges between any two 
distinct states, or $G$ could be the  network of secondary structures
for a given RNA sequence, 
whose vertices are the exponentially \cite{zukerSankoff}
many secondary structures
of the sequence, and whose edges are between any two structures $x,y$ where 
$y$ is obtained from $x$ by removing or adding a single base pair -- i.e.
$y$ is obtained by a move from $MS_1$ from $x$, or equivalently,
$x,y$ have base pair distance \cite{moulton} of $1$.  For state
$x \in X$, let $N_x$ ambiguously denote either the number of immediate
neighbors of $x$, or the set of immediate neighbors of $x$ in graph
$G$; i.e. those $y \in X$ such that there is an edge from $x$ to $y$ 
belonging to $G$. Let
$\mathbb{M}_1$ be a discrete Markov chain with set $X$ of states, having
transition probability matrix ${\bf P} = (p_{x,y})$ given by
\begin{eqnarray}
\label{eqn:transitionProb1}
p_{x,y} = \left\{ 
\begin{array}{ll}
\frac{1}{N_x} \cdot \min\left(1,\exp(-\frac{E(y)-E(x)}{RT})\right)
&\mbox{if $y \in N_x$}\\
1 - \sum_{u \in N_x} p_{x,u} &\mbox{if $x=y$}\\
0 &\mbox{if $y \not\in N_x$, $y \ne x$.}
\end{array} \right.
\end{eqnarray}
In the context of biomolecular folding (especially in {\em Markov state models} 
\cite{Swope,Huang.psb10,Pande.m10,Bowman.pnas10,Levy.ps13}), 
the transition matrix is used to describe protein and RNA folding kinetics.
Let $\mathbbp(t)$ be the time-dependent {\em population occupancy} row
vector, where $\mathbbp_i(t)$ is the probability that the molecule is
in state $i$ at (discrete) time $t \in \mathbb{N}$ -- 
here, state $i$ represents a particular 
molecular conformation in a discretized space of all possible conformations.
Then
\begin{eqnarray}
\label{eqn:kineticsMarkovChain}
\mathbbp(t) &=& \mathbbp(0) \cdot\mathbbP^t
\end{eqnarray}

Let $\mathbb{M}_2$ be a continuous Markov process with set $X$ of states,
having rate matrix ${\bf Q} = (q_{x,y})$ given by
\begin{eqnarray}
\label{eqn:transitionProb2}
q_{x,y} = \left\{ 
\begin{array}{ll}
\min\left(1,\exp(-\frac{E(y)-E(x)}{RT})\right)
&\mbox{if $y \in N_x$}\\
- \sum_{u \in N_x} q_{x,u} &\mbox{if $x=y$}\\
0 &\mbox{if $y \not\in N_x$, $y \ne x$}.
\end{array} \right.
\end{eqnarray}
In the context of biomolecular folding, the rate matrix is usually
used in the context of the {\em master equation}, also called the
{\em Kolmogorov forward equation}, defined as follows. As before,
if $\mathbbp(t)$ denotes the time-dependent {\em population occupancy} row
vector, where $\mathbbp_i(t)$ is the probability that the molecule is
in state $i$ at (continuous) time $t \in \mathbb{R}$, then the master equation
is the following system of ordinary differential equations
\begin{eqnarray*}
\frac{d \mathbbp_1(t)}{dt} &=&\sum_{j \ne 1} \left( q_{j,1} \mathbbp_j(t) - 
q_{1,j} \mathbbp_1(t)
\right)\\
\frac{d \mathbbp_2(t)}{dt} &=&\sum_{j \ne 2} \left( q_{j,2} \mathbbp_j(t) - 
q_{2,j} \mathbbp_2(t)
\right)\\
\cdots & &\\
\frac{d \mathbbp_n(t)}{dt} &=&\sum_{j \ne n} \left( q_{j,n} \mathbbp_j(t) - 
q_{n,j} \mathbbp_n(t)
\right)\\
\end{eqnarray*}
Since $q_{i,i} = - \sum_{j \ne i} q_{i,j}$, this set of
equations is equivalent to the matrix differential equation
\begin{eqnarray}
\label{eqn:populationOccupancyEquation}
\frac{d \mathbbp(t)}{dt} &=& \mathbbp(t) \cdot \mathbbQ 
\end{eqnarray}

Let $\mathcal{M_J}$ denote the {\em embedded Markov chain}, also called
{\em jump chain}, corresponding to the Markov process $\mathbb{M}_2$,
having state space $X$ and transition probability matrix $S$ given by
\begin{eqnarray}
\label{eqn:jumpChain0}
s_{x,y} = \left\{
\begin{array}{ll}
\frac{q_{x,y}}{\sum_{z \ne x} q_{x,z}} 
&\mbox{if $y \in N_x$}\\
0 &\mbox{if $x=y$ or $y \not\in N_x$}.
\end{array} \right.
\end{eqnarray}
The {\em stationary} probability distribution
${\bf p}^* = (p^*_1,\ldots,p^*_n)$ [resp. 
${\bf q}^* = (q^*_1,\ldots,q^*_n)$]
for a Markov chain [resp. Markov process]
with transition probability matrix $P=(p_{x,y})$ [resp. rate matrix
$Q=(q_{x,y})$] must satisfy
${\bf p}^* {\bf P} = {\bf p}^*$ [resp.  ${\bf q}^* {\bf Q} = {\bf 0}$].
It is well-known that
the stationary distribution exists and is unique for any finite, irreducible,
aperiodic Markov chain \cite{cloteBackofen}.
Detailed balance is said to hold, if for all states $x,y \in X$,
$p^*_x p_{x,y} = p^*_y p_{y,x}$  [resp.  $q^*_x q_{x,y} = q^*_y q_{y,x}$].

Define the Boltzmann probabilities
\begin{eqnarray}
\label{eqn:stationaryBoltzmann}
\pi_{x} = \frac{\exp(-E(x)/RT)}{Z}
\end{eqnarray}
where the partition function $Z$ is defined by
$Z = \sum_{k=1}^{n} \exp(-E(k)/RT)$.
It is trivial to verify that detailed balance holds for the rate
matrix $Q$ defined in equation~(\ref{eqn:transitionProb2}). From detailed
balance, it is easy to verify that the Boltzmann distribution is the
stationary distribution for the Markov process defined by 
equation~(\ref{eqn:transitionProb2}); indeed, if $x,y$ are neighbors
and $E(y)>E(x)$, then $q_{x,y} = \exp(-(E(y)-E(x))/RT)$ and
$\pi_x  q_{x,y} = \exp(-E(y)/RT) = \pi_y q_{y,x}$. However, as shown in
the Appendix, the Markov chain $\mathbb{M}_1$ does not generally
satisfy detailed balance for the network of RNA secondary structures.


Using Hasting's trick \cite{hastingsTrick}, where in equation 
(\ref{eqn:transitionProb1}) $p_{x,y}$ is redefined by
$p_{x,y} = \frac{1}{N_x} \cdot \min\left(1,\exp(-\frac{E(y)-E(x)}{RT}\cdot 
\frac{N_x}{N_y})\right)$ for $y \in N_x$, we obtain a Markov chain satisfying
detailed balance. Moreover the stationary distribution for the Hastings
Markov chain is easily shown to be the Boltzmann distribution from 
equation~(\ref{eqn:stationaryBoltzmann}). Nevertheless, this modification
comes at a computational cost, since $N_y$ must be determined for each
neighbor $y$ of $x$. It is perhaps for this reason that Hasting's trick is
not used in some kinetics simulations from the literature, as
in the Monte Carlo time-driven algorithm for protein
folding \cite{Sali.n94} and in Markov state models defined from
long molecular dynamics (MD) trajectories \cite{Bowman.pnas10}.

\section{Monte Carlo and Gillespie algorithms}
\label{section:HTandET}

In this section, we explore the relation between MFPT for a discrete time 
Markov chain with transition probability matrix (\ref{eqn:transitionProb1}) 
and that of the continuous time Markov process with
rate matrix (\ref{eqn:transitionProb2}).  Consider 
Algorithms~\ref{algo:1}, \ref{algo:2}, \ref{algo:3} and \ref{algo:4}.
The MFPT of the Markov chain $\mathbb{M}_1$ can be
approximated by taking the average over many
runs of Algorithm~\ref{algo:1}, as done to show that proteins fold quickly exactly
when there is a large energy gap between the energy of the native state and
that of the next lowest energy misfolded state \cite{Sali.n94}.
Time-driven Monte Carlo Algorithm \ref{algo:1} 
is equivalent to event-driven Monte
Carlo Algorithm \ref{algo:2}, since the probability of leaving state $x$ at any given
time is geometrically distributed with success probability $\Phi_1(x)$, the
probability of leaving state $x$.
By replacing the geometric distribution by the continuous exponential
distribution with the same mean, Monte Carlo Algorithms 2 and 3 are equivalent
(see Figure~\ref{fig:waitingTimeHistogramGeomertricAndExponentialDistribution}
for a comparison of relative histograms of samples from geometric
and exponential distributions having the same mean).
For a given state $x$, the {\em probability} $\Phi_1(x)$ of leaving state $x$ is
equal to the reciprocal of the number of neighbors $N_x$ of $x$ times
the {\em flux} $\Phi_2(x)$ of leaving $x$. 
Nevertheless the {\em jump probability} $s_{x,y}$ is identical in Algorithms
2,3,4 since $p_{x,y}/\Phi_1(x)$ is equal to $q_{x,y}/\Phi_2(x)$. It follows
that Algorithms 1,2,3,4 have identical probabilities of visiting exactly the
same states in a trajectory, the only difference being that the expected waiting
time in any state $x$ differs by the factor $N_x$ of the number of neighbors
of $x$.

The previous discussion shows that the only difference between the
MFPT of a discrete Markov chain and that of the related continuous time Markov
process lies in the fact that the time increment $\Delta t$ in the former
is sampled from a geometric distribution with mean
$N_x/\sum_{y \in N_x} \min\left(1,\exp(-\frac{E(y)-E(x)}{RT})\right)$
while the time increment $\Delta t$ in the latter
is sampled from a exponential distribution with mean
$1/\sum_{y \in N_x} \cdot$ $\min\left(1,\exp(-\frac{E(y)-E(x)}{RT})\right)$.
Hence, if the value $N_x$ equals a fixed
constant $N$ for each state $x$, then the Markov chain MFPT equals 
$N$ times the Markov process MFPT, as shown in the following example.
\medskip

\noindent
\begin{example}
\label{example1}
Let $X = \{ 0,1,\ldots,2n-1 \}$ be the set of $2n$ states, 
where the energy of state $k$ is given by
\begin{eqnarray}
\label{eqn:energyOfStatesToyMarkovChain}
E(k) = \left\{  \begin{array}{ll}
-3 &\mbox{if $k=0$}\\
-2 &\mbox{if $k>0$ and $k$ is even}\\
-1 &\mbox{if $k$ is odd.}\\
\end{array} \right.
\end{eqnarray}
The left panel of
Figure~\ref{fig:comparisonFourMCalgorithms} depicts the $6$-node
network $X$, where $n=3$.
Let $Adj(i,j)$ be the indicator function for whether $i,j$ are adjacent in
a circle of size $2n$; i.e.  
\begin{eqnarray}
\label{eqn:adjacencyOfCircle}
Adj(i,j) &=& I[ j \equiv  (i \pm 1) \bmod 2n].
\end{eqnarray}
Since the underlying graph is $2$-regular, $N_x=2$ for each state $x$, and
the transition probability matrix of $\mathbb{M}_1$ is defined by
\begin{eqnarray}
\label{fig:transitionProbMatrixToyMarkovChain}
p_{i,j} = \left\{  \begin{array}{ll}
0.5 \min(1,\exp(-(E(j)-E(i)))) &\mbox{if $Adj(i,j)$}\\
1-p_{i,i+1 \bmod 2n}-p_{i,i-1 \bmod 2n} &\mbox{if $i=j$}\\
0 &\mbox{otherwise}\\
\end{array} \right.
\end{eqnarray}
while the rate matrix of $\mathbb{M}_2$ is defined by
\begin{eqnarray}
\label{fig:rateMatrixToyMarkovChain}
q_{i,j} = \left\{  \begin{array}{ll}
\min(1,\exp(-(E(j)-E(i)))) &\mbox{if $Adj(i,j)$}\\
-p_{i,i+1 \bmod 2n}-p_{i,i-1 \bmod 2n} &\mbox{if $i=j$}\\
0 &\mbox{otherwise}\\
\end{array} \right.
\end{eqnarray}
\end{example}

\subsubsection*{Relation between $\mathcal{M}_1$ and $\mathcal{M}_2$}

Given that the definition of the
transition probability matrix ${\bf P} = (p_{x,y})$ 
of the Markov chain $\mathcal{M}_1$, as given in 
equation~(\ref{eqn:transitionProb1}), is similar to that of the
rate matrix ${\bf Q} = (q_{x,y})$ of the Markov process $\mathcal{M}_2$,
as given in equation~(\ref{eqn:transitionProb2}), it is natural to
ask what the relation is between ${\bf P}$ and ${\bf Q}$. The answer is
that there is no relation. For the Markov chain $\mathcal{M}_1$,
equation~(\ref{eqn:kineticsMarkovChain}) states that the population
occupancy vector at time $t=1$ satisfies
$\mathbbp(1) = \mathbbp(0) \cdot\mathbbP$.
In contrast, for the Markov process $\mathcal{M}_2$,
equation~(\ref{eqn:populationOccupancyEquation}) states that the population
occupancy vector at time $t=1$ satisfies
$\mathbbp(1) = \mathbbp(0) \cdot \exp(\mathbbQ)$,
since the solution of the master equation
(\ref{eqn:populationOccupancyEquation}) is
\begin{eqnarray}
\label{eqn:solutionMasterEquation}
\mathbbp(t) &=& \mathbbp(0) \cdot \exp(\mathbbQ \cdot t)
\end{eqnarray}
For instance, if we set $n=2$ in Example~\ref{example1}, then the transition
probability matrix is
\begin{align*}
\mathbbP &= \left( \begin{array}{llllllll}
0.8647& &0.0677& &0.0000& &0.0677\\ 
0.5000& &0.0000& &0.5000& &0.0000\\ 
0.0000& &0.1839& &0.6321& &0.1839\\ 
0.5000& &0.0000& &0.5000& &0.0000\\ 
\end{array} \right)
\end{align*}
while the rate matrix is
\begin{align*}
\mathbbQ &= \left( \begin{array}{llllllll}
-0.2707& &0.1353& &0.0000& &0.1353\\ 
1.0000& &-2.0000& &1.0000& &0.0000\\ 
0.0000& &0.3679& &-0.7358& &0.3679\\ 
1.0000& &0.0000& &1.0000& &-2.0000\\ 
\end{array} \right)
\end{align*}
and the exponential of the rate matrix is
\begin{align*}
\exp(\mathbbQ) &= \left( \begin{array}{llllllll}
0.8299& &0.0566& &0.0569& &0.0566\\ 
0.4183& &0.1983& &0.3205& &0.0629\\ 
0.1547& &0.1179& &0.6095& &0.1179\\ 
0.4183& &0.0629& &0.3205& &0.1983\\ 
\end{array} \right)
\end{align*}
Clearly, the value $\mathbbp(1) = \mathbbp(0) \cdot\mathbbP$ obtained by
equation~(\ref{eqn:kineticsMarkovChain}) is unequal to the value
$\mathbbp(1) = \mathbbp(0) \cdot \exp(\mathbbQ)$ obtained by
equation~(\ref{eqn:populationOccupancyEquation}).

\subsubsection*{Mean first passage time for Example~\ref{example1}}

Using the 20 state Markov chain obtained in Example~\ref{example1} by
setting $n=10$,
the time to reach the minimum energy state $0$ was computed for the
(1) time-driven Monte Carlo Algorithm~\ref{algo:1},
(2) event-driven Monte Carlo Algorithm~\ref{algo:2}
with geometrically distributed waiting times,
(3) event-driven Monte Carlo Algorithm~\ref{algo:3}
with exponentially distributed waiting times, 
(4) Gillespie's Algorithm~\ref{algo:4}.  Times for Algorithms 
\ref{algo:1}, \ref{algo:2} and \ref{algo:3}
were divided by $2$, since
the number $N_x$ of neighbors of each state $x$ is $2$, while times for
Algorithm~\ref{algo:4} were reported as computed. 
The average time was taken over 10,000 separate
runs of each algorithm, and then histograms were produced for 1000 repetitions
of each of the 10,000 runs. It follows by the central limit theorem that
each histogram is approximately normal; moreover, the mean and standard
deviation for each histogram is reported as follows:
(1) $\mu=92.94$, $\sigma=0.76$,
(2) $\mu=92.97$, $\sigma=0.75$,
(3) $\mu=92.96$, $\sigma=0.75$,
(4) $\mu=92.92$, $\sigma=0.78$. 
The right panel of Figure~\ref{fig:comparisonFourMCalgorithms}
displays superimposed histograms of for the expected number 
of moves obtained on average for Algorithms 1,2,3,4
for a state set $X$ with $20$ states $0,1,\ldots,19$.

\section{Results}
\label{section:results}
\subsection{Expected trajectory time}


Throughout this section, we assume that $\mathbb{M}_1$ [resp. $\mathbb{M}_2$] is
a finite, irreducible, aperiodic Markov chain [resp. Markov process] whose
transition probabilities $p_{x,y}$ [resp.  transition rates $q_{x,y}$]
satisfy equation~(\ref{eqn:transitionProb1})
[resp. equation~(\ref{eqn:transitionProb2})]. Moreover, we assume that
$\mathbb{M}_1$ and $\mathbb{M}_2$ have the same underlying set 
$X=\{1,\ldots,n\}$ of states, and that each state $x \in X$
is labeled by the same energy $E(x)$ in both $\mathbb{M}_1$ and $\mathbb{M}_2$.

The main result of this section is that asymptotically, for large values of $K$,
the expected time taken by a $K$-transition 
trajectory\footnote{Each step of Algorithms 
\ref{algo:2}, \ref{algo:3} and \ref{algo:4} involves a transition from
a current state to a distinct state; however, each step of the time-driven 
Monte Carlo Algorithm~\ref{algo:1} does not necessarily involve a transition 
to a new state,
especially if the current state $x$ has low energy, so that $p_{x,x}$ may be
large. For this reason, we use the term $K$-transition trajectory.}
for each of the Monte Carlo
Algorithms \ref{algo:1}, \ref{algo:2}, \ref{algo:3} is equal to $\mu$ 
multiplied by the expected time for a $K$-transition trajectory of the 
Gillespie Algorithm~\ref{algo:4}, where $\mu$ denotes the
expected degree as defined in equation~(\ref{eqn:networkDegree}); i.e. the
expected number of neighbors, given by
\[
\sum_{x \in X} \frac{\exp(-E(x)/RT)}{Z} \cdot N_x
\]

Let Algorithm 1$^{\dag}$, 2$^{\dag}$, 3$^{\dag}$, 4$^{\dag}$ respectively
denote Algorithm 1,2,3,4 where line 3 of each algorithm is replaced by
the line ``{\bf \tt while {\sc true}}''; i.e. there is no condition on the
while loop, so the algorithms do not terminate. 
In Theorem~\ref{thm1}, 
we establish that for any given RNA sequence, as the number 
$K$ of trajectory steps approaches infinity, the trajectory time of 
Algorithm 3$^{\dag}$ equals that of Algorithm 4$^{\dag}$  multiplied by
the expected network degree $\mu$.  In the previous section, we showed
that Algorithms 1,2,3 are equivalent; it follows that
Algorithms 1$^{\dag}$, 2$^{\dag}$, 3$^{\dag}$ are also equivalent. Thus
Theorem~\ref{thm1} directly relates (asymptotic)
{\em trajectory time} of Monte Carlo time-driven and event-driven
algorithms with that of Gillespie's algorithm.
In order to give a formal statement of this result, 
we need to provide some definitions. 

%


\begin{definition}[$K$-step trajectory and expected trajectory time]
\label{def:KstepTrajectoryAndExpectedTrajectoryTime}
A $K$-step trajectory (sometimes called $K$-transition trajectory)
of Monte Carlo Algorithm 3$^{\dag}$ [resp. Gillespie Algorithm 4$^{\dag}$] 
is a sequence $x_0,x_1,\ldots,x_K$ of states in $\{1,\ldots,n\}$, 
where $x_i \ne x_{i+1}$ for $i=0,\ldots,K$, although 
$x_i = x_j$ can occur for $|j-i| \geq 2$. Let
$\tau^K_3$ [resp. $\tau^K_4$] denote the random variable whose value is
the sum of the time increments $\Delta t$ in line 11 of 
Algorithm 3$^{\dag}$ [resp. Algorithm 4$^{\dag}$]. 
Denote the {\em expected time} for $K$ steps of Algorithm 3$^{\dag}$ 
[resp. Algorithm 4$^{\dag}$] by $E[\tau^K_3]$ or
$\langle \tau^K_3 \rangle$ [resp. $E[\tau^K_4]$ or
$\langle \tau^K_4 \rangle$],
where the expectation is taken over all stochastically 
sampled $K$-step trajectories (lines 13-18) and sampled time increments 
(line 11) of Algorithm 3$^{\dag}$ [resp. Algorithm 4$^{\dag}$].
\end{definition}

\noindent
\begin{theorem}
\label{thm1}
Let $\mu = \sum_{x \in X} \frac{\exp(-E(x)/RT)}{Z} \cdot N_x$
denote the Boltzmann expected network degree, where
$N_x$ denotes the degree of state $x$. Then
$\lim_{K \rightarrow \infty} 
\frac{E[\tau^K_1]}{E[\tau^K_2]} = \mu$.
\end{theorem}
\medskip

\noindent
{\bf \sc Proof:} For each state $x \in X$, define the 
Boltzmann probability $\pi_{x}$ is defined by 
$\pi_{x} = \frac{\exp(-E(x)/RT)}{Z}$, where
the partition function $Z$ is defined by
$\sum\limits_{x \in X} \exp(-E(x)/RT)$.
For state $x \in X$, define the {\em probability} of leaving $x$
[resp. {\em flux} out of $x$], denoted $\Phi_1(x)$ [resp. $\Phi_2(x)$] by
\begin{eqnarray}
\label{eqn:rateFlux}
\Phi_1(x)  = \sum_{y \ne x} p_{x,y} = \frac{\Phi_2(x)}{N_x}\\
\label{eqn:probabilityFlux}
\Phi_2(x)  = \sum_{y \ne x} q_{x,y} = N_x \cdot \Phi_1(x)
\end{eqnarray}
Recall that the transition probability matrix $S$ 
of the {\em embedded chain}, also called {\em jump chain},
$\mathbb{M}_J$ is defined in equation~(\ref{eqn:jumpChain0}), and
hence satisfies
\begin{eqnarray}
\label{eqn:jumpChain}
s_{x,y} = \left\{
\begin{array}{ll}
\frac{q_{x,y}}{\Phi_2(x)} 
= \frac{p_{x,y}}{\Phi_1(x)} 
&\mbox{if $y \in N_x$}\\
0 &\mbox{if $x=y$ or $y \not\in N_x$}.
\end{array} \right.
\end{eqnarray}
A $K$-step trajectory of either Algorithm~3$^{\dag}$ or 4$^{\dag}$ corresponds
to a random walk on the jump chain with states $1,\ldots,n$ and transition
matrix $S$ as defined in equation~(\ref{eqn:jumpChain}).
Since the rate matrix
$Q$, defined in equation~(\ref{eqn:transitionProb2}),
for the Markov process $\mathbb{M}_2$ corresponding to Algorithm~4$^{\dag}$
trivially satisfies detailed balance, so does the transition matrix $S$ 
for the jump chain, defined in equation~(\ref{eqn:jumpChain}).  
\medskip

\noindent
{\sc Claim:} Define the row vector 
$s^* = (s^*_1,\ldots,s^*_n)$ by
\begin{eqnarray}
\label{eqn:stationaryJumpChain}
s^*_x &=&
\frac{q^*_x \Phi_2(x)}{\sum_{y \in \mathbb{SS}({\bf a})} q^*_y \Phi_2(y)}.
\end{eqnarray}
Then $s^* S = s^*$, hence $s^*$ is the (unique) stationary distribution for
the jump chain $\mathbb{M}_J$.
\smallskip

\noindent
{\sc Proof:}
For fixed state $y \in \{1,\ldots,n\}$, the $y$th coordinate of the row vector
$s^* S$ satisfies
\begin{eqnarray*}
(s^* S)_y &=&
\sum_{x=1}^n  s^*_x \cdot s_{x,y} = \sum_{x \ne y}  s^*_x \cdot s_{x,y}  \\
&=&
\sum_{x \ne y}
\left( \frac{q^*_x \Phi_2(x)}{\sum_{z \in \mathbb{SS}({\bf a})} q^*_z \Phi_2(z)}
\right) \cdot \frac{q_{x,y}}{\Phi_2(x)}\\
&=&
\sum_{x \ne y}
\frac{q^*_x q_{x,y} }{\sum_{z \in \mathbb{SS}({\bf a})} q^*_z \Phi_2(z)} 
=
\sum_{x \ne y}
\frac{q^*_y q_{y,x} }{\sum_{z \in \mathbb{SS}({\bf a})} q^*_z \Phi_2(z)} \\
&=&
q^*_y \cdot \left( \sum_{x \ne y} q_{y,x} \right) 
\frac{1}{\sum_{z \in \mathbb{SS}({\bf a})} q^*_z \Phi_2(z)}  \\
&=&
\frac{q^*_y \Phi_2(y)}{\sum_{z \in \mathbb{SS}({\bf a})} q^*_z \Phi_2(z)} 
=
s^*_y
\end{eqnarray*}
Note that line 2 follows by definition of $s^*_x$ and $s_{x,y}$;
line 3 follows by detailed balance of $q^*_x q_{x,y} = q^*_y q_{y,x}$;
line 4 follows by factoring out terms that do not depend on $x$, and
line 5 follows by the definition of $\Phi_2(y)$.

An equivalent, more intuitive statement of 
equation~(\ref{eqn:stationaryJumpChain}) is that $s^*_y$ equals the
the Boltzmann probability $q^*_y$ of state $y$, times the flux
$\Phi_2(y)$ out of $y$, divided by the {\em expected flux} with respect to
the Boltzmann distribution; i.e. $s^*_y$ is the ratio of the Boltzmann
weighted flux out of $y$ with respect to the expected flux out of any
state, where expectation is taken over all states. This proves
the claim.

If the number $K$ of steps in the trajectory is large, then the expected
number of occurrences of each state $x$ is $s^*_x K$. The expected time to
leave state $x$ in Algorithm~3$^{\dag}$ [resp. Algorithm~4$^{\dag}$] is
$\frac{1}{\Phi_1(x)}$ [resp.  $\frac{1}{\Phi_2(x)}$], since waiting times
are exponentially distributed. It follows that the ratio of
the expected time for a $K$-step trajectory of Algorithm~3$^{\dag}$  divided by
the expected time for a $K$-step trajectory of Algorithm~4$^{\dag}$ equals: 
\begin{eqnarray*}
\frac{E[\tau^K_1]}{E[\tau^K_2]}  &=&
 \frac{\sum_{x=1}^n \frac{s^*_x K}{\Phi_1(x)}}
{\sum_{x=1}^n \frac{s^*_x K}{\Phi_2(x)}} =
\frac{\sum_{x=1}^n \frac{s^*_x N_x}{\Phi_2(x)}}
{\sum_{x=1}^n \frac{s^*_x}{\Phi_2(x)}}\\
&=&
\frac{ \sum_{x=1}^n \frac{q^*_x \Phi_2(x)}{\sum_z q^*_z \Phi_2(z)} \cdot
\frac{N_x}{\Phi_2(x)}}
{ \sum_{x=1}^n \frac{q^*_x \Phi_2(x)}{\sum_z q^*_z \Phi_2(z)} \cdot
\frac{1}{\Phi_2(x)}} \\
&=&
\frac{ \sum_{x=1}^n q^*_x  N_x }
{ \sum_{x=1}^n q^*_x  } =
\frac{ \sum_{x=1}^n \pi_x  N_x }
{ 1 } \\
&=&
\sum_{x=1}^n \frac{\exp(-E(x)/RT) \cdot N_x}{Z} = \mu({\bf a}).
\end{eqnarray*}
This completes the proof of Theorem~\ref{thm1}.  $\blacksquare$
\medskip

\noindent
\begin{corollary}
\label{cor1}
The same proof shows that mean recurrence time for
Markov chain $\mathbb{M}_1$ is equal to $\mu$ times the mean recurrence time
for Markov process $\mathbb{M}_2$. (See \cite{feller1} for the definition
of mean recurrence time.)
\end{corollary}

\subsection{Illustrative examples from RNA}

For a fixed RNA sequence ${\bf a}=a_1,\ldots,a_m$, denote the
set of all secondary structures for ${\bf a}$ by $\mathbb{SS}({\bf a})$.
Define the Markov chain $\mathbb{M}_1$ [resp. Markov process $\mathbb{M}_2$]
for the set $\mathbb{SS}({\bf a})$ of states, where transitions 
$s \rightarrow t$ occur between secondary structures $s,t$ that differ by base
pair distance of $1$. $\mathbb{M}_1$ [resp. $\mathbb{M}_2$] has
probability matrix $P=(p_{x,y})$ [resp. rate matrix $Q=(q_{x,y})$]
defined by equation~(\ref{eqn:transitionProb1})
[resp. equation (\ref{eqn:transitionProb2})], where $E(x)$ denotes the
free energy of secondary structure $x$ with respect to the Turner energy
model \cite{Turner.nar10}. Theorem~\ref{thm1}
then implies that for trajectories of
length $K$, for large $K$, the expected trajectory time
$E[\tau^K_1]$ for each of Algorithm 1$^{\dag}$, Algorithm 2$^{\dag}$, and
Algorithm 3$^{\dag}$ is approximately $\mu$ multiplied by the trajectory time
$E[\tau^K_2]$ for Algorithm 4$^{\dag}$, where $\mu$ is the expected
degree for the network $\mathbb{SS}({\bf a})$ of secondary structures for
RNA sequence ${\bf a} = a_1,\ldots,a_n$.

Although the set $\mathbb{SS}({\bf a})$ is generally of size exponential
in $n$ (see \cite{steinWaterman,Fusy.jmb12}), we recently described a cubic 
time dynamic programming algorithm \cite{cloteJCC2015} to
compute the {\em expected network degree} of $\mathbb{S}({\bf a})$
\[
\mu({\bf a}) = \sum\limits_{x \in \mathbb{SS}({\bf a})}
\frac{\exp(-E(x)/RT)}{Z} \cdot N_x
\]
where $N_x$ is the set of secondary structures $y$ having base pair distance
$1$ to $x$.  

We illustrate Theorem~\ref{thm1} in the following computational experiments.  
For a given RNA sequence ${\bf a}$, let $x_0$ denote the empty secondary
structure containing no base pairs. If algorithms
C${^\dag}$ and D${^\dag}$ begin from the same initial state $x_0$ and
use the same pseudo-random number generator seed, then the algorithms
are said to be {\em synchronized}, or to generate {\em synchronized
trajectories}; if no random seed is set, then the algorithms are said to be 
{\em unsynchronized}. 
If Algorithms 3${^\dag}$ and 4${^\dag}$ are synchronized, then clearly
each algorithm visits exactly the
same states in the same order.

The following computational experiment is described. Given an RNA sequence,
${\bf a} = a_1,\ldots,a_n$, we determine the minimum free energy (MFE)
structure $s^*$ by Zuker's algorithm \cite{zukerStiegler} as implemented in
Vienna RNA Package \cite{Lorenz.amb11}.  For the 32 nt fruA SECIS element
{\tt CCUCGAGGGG AACCCGAAAG GGACCCGAGA GG}, 
for which the target minimum free energy
structure is reached generally in less than one thousand moves,
we analyzed a trajectory of one million moves.  For {\em synchronized} runs of
Algorithm~3$^{\dag}$ [resp. Algorithm~4$^{\dag}$] the trajectory time
was approximately $98.16168 \cdot 10^9$ 
[resp. $9.82994 \cdot 10^9$] 
expected number of neighbors
$\mu({\bf a}) = 10.00146$, and the ratio of Algorithm~3$^{\dag}$
trajectory time divided by $\mu({\bf a})$
is $9.81474 \cdot 10^9$ -- close to the Algorithm~4$^{\dag}$ trajectory
time. 
For {\em nonsynchronized} runs of
Algorithm~3$^{\dag}$ [resp. Algorithm~4$^{\dag}$] the trajectory time
was approximately $98.03194 \cdot 10^9$ 
[reps. $9.78867 \cdot 10^9$], 
and the ratio of Algorthm C$^{\dag}$ trajectory time divided by
$\mu({\bf a})$
is $9.80177 \cdot 10^9$ -- very close to Algorithm~4$^{\dag}$ trajectory
time. 
Similar validation of Theorem~\ref{thm1} was shown in
second computational experiment, where a 
500 thousand step trajectory was
generated for the
76 nt ala-tRNA from {\em Mycoplasma mycoides} with Sprinzl ID RA1180 
(tRNAdb ID tdbR00000006) \cite{Juhling.nar09} using Algorithms
C$^{\dag}$ and D$^{\dag}$. A final illustration of  Theorem~\ref{thm1}
is given in the right panel of 
Figure~\ref{fig:comparisonFourMCalgorithms}, which displays a scatter plot,
for both synchronized and unsynchronized runs of one million steps for the
Monte Carlo Algorithm~3$^{\dag}$ and the Gillespie Algorithm~4$^{\dag}$ for
forty 20 nt randomly generated RNA sequences, each of whose MFE
secondary structure has free energy $E < -2.5$ kcal/mol, each 
having at most 2500 secondary structures, and each
with expected compositional frequency of 0.25 for each of A,C,G,U.

%
%


Figure~\ref{fig:RNAsuboptNumNbors} shows the distribution for the number of
neighbors 
the 76 nt transfer RNA (RA1180 from tRNAdb 2009 \cite{Juhling.nar09}), 
with mean $29.29$ and standard deviation $4.14$, as well as that 
for the 56 nt spliced leader RNA from {\em Leptomonas collosoma}
(a known conformational switch), with mean
$70.05$ and standard deviation of $33.84$.  Although
we can exactly compute the expected network degree in
seconds \cite{cloteJCC2015},
the only current method to obtain an approximation of the distribution is
by sampling using {\tt RNAsubopt} \cite{Wuchty.b99} from the
Vienna RNA Package \cite{Lorenz.amb11}, a computation requiring many hours.
It follows that the RNA examples illustrating Theorem~\ref{thm1} 
could not have been
given without use of the dynamic programming algorithm to compute expected
network degree \cite{cloteJCC2015}.
\medskip

\begin{remark}
\label{remark1}
In our examples of RNA secondary structure trajectory
time, we considered elementary step transitions between
secondary structures $x,y$, in which $y$ is obtained from $x$ by addition or
removal of a single base pair. However, Theorem~\ref{thm1} 
is equally applicable
for more general move sets between RNA secondary structures, including
{\em shift moves} \cite{flamm} and even the addition and removal of entire
stems \cite{Isambert.pnas00}. In each of these cases, it is easily established
that the underlying Markov chain [resp. Markov process] is finite, irreducible
and aperiodic, so that Theorem~\ref{thm1} applies.
\end{remark}

\begin{remark}
\label{remark2}
The program {\tt Kinfold} \cite{flamm} is an implementation of the Gillespie
algorithm, which supports both move set $MS_1$, consisting of adding or
removing a single base pair, and move set $MS_2$, consisting of adding,
removing, or {\em shifting} a base pair. Since we found it difficult to
modify the source code of {\tt Kinfold} to implement the Monte Carlo algorithm,
we implemented the Monte Carlo and Gillespie algorithms for RNA secondary
structure networks in a C-program {\tt mc.c}, available at
\url{http://bioinformatics.bc.edu/clotelab/RNAexpNumNbors/}.
In \cite{Clote.po15}, we generalized
the algorithm from \cite{cloteJCC2015} to efficiently 
compute the expected network
degree for RNA secondary structure networks with respect to move set
$MS_2$. At the present time, however, our program {\tt mc.c} does not
support shift moves.
\end{remark}

\subsection{Monte Carlo and Gillespie MFPT}

Here, we show that unless the network is regular (each node has the
same degree), there is no simple analogue of Theorem~\ref{thm1}
to relate the mean first passage time (MFPT) for the Monte Carlo Algorithm~3
and the Gillespie Algorithm~4 for RNA secondary structure folding kinetics.  
Algorithms 3$^{\dag}$ and 4$^{\dag}$ differ
from Algorithms 3 and 4, in that the former algorithms do not terminate
computation upon visiting the minimum free energy structure $x_{\infty}$.
This modification allowed us to establish Theorem~\ref{thm1}.

In Example~\ref{example1} from Section~\ref{section:HTandET}, 
we described the computation of mean first passage time
(MFPT) using each of Algorithms 1,2,3,4 for a Markov chain $\mathbb{M}_1$
[resp. Markov process $\mathbb{M}_2$], having states
$2n$ states $0,1,\ldots,2n-1$, where energy $E(x)$ of state $x$ is defined
in equation~(\ref{eqn:energyOfStatesToyMarkovChain}), and transition probability
$p_{x,y}$ of moving from state $x$ to state $y \in \{ x-1 \bmod 2n, x+1
\bmod 2n\}$  is defined in 
equation~(\ref{fig:transitionProbMatrixToyMarkovChain}).
Since the underlying graph is $2$-regular, 
Figure~\ref{fig:comparisonFourMCalgorithms} indeed shows
that MFPT for each of 
Monte Carlo Algorithms 1,2,3 is twice the MFPT of Gillespie Algorithm 4.

By slightly modifying the topology of Example~\ref{example1}, 
we obtain a non-regular network with states $0,\ldots,2n-1$, 
for which all states have degree $2$, with the exception of
the (terminal) minimum energy state $0$ and the last state $2n-1$. For
this example, described in the following,
there appears to be no simple relation between the MFPT of Monte
Carlo Algorithms 1,2,3 with that of Gillespie Algorithm 4.
\medskip

\noindent
\begin{example}
\label{example2}
Define Markov chain $\mathbb{M}_1$ [resp. Markov process
$\mathbb{M}_2$] as in Example~\ref{example1}, 
with the sole exception that state $0$ is
no longer connected to state $2n-1$; i.e. the indicator
function $Adj(i,j)$ for whether $i,j$ are adjacent is redefined by
\begin{eqnarray}
\label{eqn:adjacencyOfCircleLinearized}
Adj(i,j) &=& \left\{ \begin{array}{ll}
1 &\mbox{if $0 \leq i < 2n-1$ and $j=i+1$}\\
1 &\mbox{if $0 \leq j < 2n-1$ and $i=j+1$}\\
0 &\mbox{otherwise}
\end{array} \right.
\end{eqnarray}
Transition probabilities [resp. rates] are defined by 
equation~(\ref{eqn:transitionProb1})
[resp. equation~(\ref{eqn:transitionProb2})] where we set $RT=1$, hence
\begin{eqnarray}
\label{fig:transitionProbMatrixToyMarkovChainBis}
p_{i,j} = \left\{  \begin{array}{ll}
0.5 \cdot \min(1,\exp(-(E(j)-E(i)))) 
  &\mbox{if $Adj(i,j), 1 \leq i,j < 2n-1$}\\
\min(1,\exp(-(E(j)-E(i)))) 
  &\mbox{if $i=0,j=1$ or $i=2n-1,j=2n-2$}\\
1-p_{i,(i+1 \bmod 2n)}-p_{i,(i-1 \bmod 2n)} &\mbox{if $i=j$}\\
0 &\mbox{otherwise}\\
\end{array} \right.
\end{eqnarray}
while the rate matrix of $\mathbb{M}_2$ remains unmodified, defined by
\begin{eqnarray}
\label{fig:rateMatrixToyMarkovChainBis}
q_{i,j} = \left\{  \begin{array}{ll}
\min(1,\exp(-(E(j)-E(i)))) &\mbox{if $Adj(i,j)$}\\
-p_{i,(i+1 \bmod 2n)}-p_{i,(i-1 \bmod 2n)} &\mbox{if $i=j$}\\
0 &\mbox{otherwise}\\
\end{array} \right.
\end{eqnarray}
\end{example}
Let $n=10$, so that there are $20$ states $0,\ldots,19$, and let the
initial state $x_0 = 10$, as in Example~\ref{example1}.
The time to reach the minimum energy state $0$ was computed for the
time-driven Monte Carlo Algorithm~1, event-driven Monte Carlo Algorithm~2
with geometrically distributed waiting times, event-driven Monte Carlo 
Algorithm~3 with exponentially distributed waiting times and
the Gillespie Algorithm~4.  The average time was taken over 10,000 separate
runs of each algorithm, and histograms were produced for 1000 repetitions
of each of the 10,000 runs. The mean and standard
deviation for each histogram is reported as follows:
(1) $\mu = 520.93$, $\sigma = 16.99$,
(2) $\mu = 520.52$, $\sigma=16.97$,
(3) $\mu = 520.67$, $\sigma=17.00$,
(4) $\mu = 265.21$, $\sigma=8.62$.
Figure~\ref{fig:comparisonFourMCalgorithmsBis} displays the relative
histograms for this data, where for comparison purposes, first passage
times for Gillespie Algorithm~4 are multiplied by $2$. The uniform
expected number of neighbors, or network degree, computed over the
the collection of 19 non-terminal states $x \ne 0$, is
$1.947$, 
while the Boltzmann expected number of neighbors is 
$1.710$. 
In contrast, the  ratio of the average $520.706$ of the mean first passage
times computed by Monte Carlo Algorithms 1,2,3 divided by the mean first passage
time $265.209$ for the Gillespie Algorithm 4 is equal to $1.963$. 
The p-value for the 2-tailed T-test for equality of Monte Carlo
MFTP $520.706$ with Gillespie MFPT $265.21$ times {\em uniform} network degree
$1.947$ is $1.93149 \cdot 10^{-11}$. It follows that Monte Carlo MFPT is
statistically different from Gillespie MFPT times the expected degree of
the network in the case of Example~\ref{example2}
(for either uniform or Boltzmann probability).

\subsection{Folding trajectories for the 10-mer GGGGGCCCCC}
\label{section:foldingTrajectoriesFor10mer}

Theorem~\ref{thm1} implies that IF there is {\em no termination} condition
for Algorithms 3 and 4 (obtained by replacing line 3 by 
{\bf while {\sc true}}), so that every secondary structure of the 10-mer
GGGGGCCCCC may be visited multiple times (including the MFE structure), 
THEN for a sufficiently large number of algorithm steps $K$, the
total time for the trajectory of Algorithm 3 is (asymptotically) equal
to the total time for the trajectory of Algorithm 4 times the {\em expected
network degree}. We now show that this is \underline{false} 
for mean first passage
times; i.e. if Algorithms 3 and 4 terminate upon reaching the MFE
structure, then there is no such relation between their trajectory times,
which correspond to the first passage time from the empty initial structure
to the MFE structure. Before proceeding, we need some notation.

\begin{definition}[Expected convergence time]
\label{def:MFEtrajectoryAndExpectedTrajectoryTime}
Given an RNA sequence ${\bf a} = a_1,\ldots,a_n$,
let $\tau^{\mbox{\tiny MFE}}_3({\bf a})$ [resp.  
$\tau^{\mbox{\tiny}{MFE}}_4({\bf a})$]
denote the random variable whose value is
the sum of the time increments $\Delta t$ in line 11 of 
Algorithm~3 [resp. Algorithm~4] until the algorithm converges,
where initial state $x_0$ is the empty secondary structure 
of {\bf a} and final state $x_{\infty}$ is the MFE structure of ${\bf a}$.  
Let $E[\tau^{\mbox{\tiny MFE}}_3({\bf a})]$ or 
$\langle \tau^{\mbox{\tiny MFE}}_3({\bf a}) \rangle$ 
[resp.  $E[\tau^{\mbox{\tiny MFE}}_4({\bf a})]$ or 
$\langle \tau^{\mbox{\tiny MFE}}_4({\bf a}) \rangle$]
denote the expected convergence time of Algorithm~3
[resp. Algorithn~4], or in other words, the \underline{mean first passage time}
to fold the RNA sequence {\bf a} using Monte Carlo Algorithm~3
[resp. Gillespie Algorithm~4].
\end{definition}

For the 10 nt RNA sequence GGGGGCCCCC, we ran Algorithms 3 and 4 to compute
the trajectory time for 100,000 distinct, {\em synchronized} folding 
trajectories, each trajectory starting from the empty initial structure
$x_0 = \emptyset$ and terminating upon reaching 
the minimum free energy (MFE) structure 
$x_{\infty} = \{ (1,10),(2,9),(3,8) \}$\footnote{Recall that secondary
structures may be represented as a set of base pairs.}
with dot-bracket notation
$\op \op \op \bullet \bullet \bullet \bullet \cp \cp \cp$.
Figure~\ref{fig:relativeHistogramAlgoCandDandCminusDegTimesD}a
[resp. 
Figure~\ref{fig:relativeHistogramAlgoCandDandCminusDegTimesD}b]
depicts the  relative histogram for the 100,000 trajectory times of 
Algorithm~3 [resp. 4] -- note that the first passage times
appear to be exponentially distributed, although the fit is not good.
Since the folding experiments were
synchronized, for each $i=1,\ldots,100,000$, the number of steps taken by
each of Algorithms \ref{algo:3} and \ref{algo:4} 
were identical, as were the secondary structures 
visited in each step, so that the only difference between 
Algorithms \ref{algo:3} and \ref{algo:4} 
consisted in the incremental times $\Delta t$ determined in line 11 of 
each algorithm, as well as in the total trajectory time in line 20 of each
algorithm.

\subsubsection*{Expected degree with respect to the Boltzmann distribution}

Using the software from \cite{cloteJCC2015}, we find that {\em Boltzmann}
expected number of neighbors $\langle N \rangle$, as defined in
equation~(\ref{eqn:networkDegree}), for the
set of all 62 secondary structures of the 10-mer GGGGGCCCCC is
$\langle N \rangle  = 3.031162 \approx 3.03$. 
Figure~\ref{fig:relativeHistogramAlgoCandDandCminusDegTimesD}c
depicts the  relative histogram for the pairwise differences
$T_c - \langle N \rangle \cdot T_d$. 
Let the null hypothesis $H_0$ assert that
trajectory time $T_c$ for Algorithm~3
to reach the MFE structure is equal to $\langle N \rangle$ multiplied by
the trajectory time $T_d$ for Algorithm~4
to reach the MFE structure; i.e. $H_0$ is the assertion that
$T_c = 3.031162 \cdot T_d$.

The p-value for $n=100,000$
paired values $(T_c, \langle N \rangle \cdot T_d)$ is $9.0923E-107$, since
the mean $\overline{x}_{dif}$
of paired differences $T_c- \langle N \rangle \cdot T_d$ 
satisfies
$\overline{x}_{dif} = 
1277.20$ $= \sum_{i=1}^n \frac{T_c[i]-\langle N \rangle \cdot T_d[i]}{n}$,
the standard deviation $s_{dif}$
of the paired differences satisfies $s_{dif}=18379.89$,
the test statistic $t = 21.97 = \frac{\overline{x}_{dif}}{s_{dif}/\sqrt{n}}$,
and the number $df$ of degrees of freedom $n-1=99,999$.

For $\alpha=0.05$, to compute the $1-\alpha = 95\%$ confidence interval for 
$T_c-\langle N \rangle \cdot T_d$, we determine the margin of error 
$E = t_{\alpha/2} \cdot \frac{s_{dif}}{\sqrt{n}} \approx
1.645 \cdot 58.12 = 95.60$, yielding a confidence interval of
$(\overline{x}_{dif}-E, \overline{x}_{dif}+E) = (1181.59,1372.80)$.
Under the null hypothesis,
$\overline{x}_{dif}$ is asserted to be $0$, which does not belong
to the confidence interval. Even repeating the computation with
$\alpha=10^{-100}$, we still observe that $0$ does not belong to the
corresponding confidence interval $(39.33,2515.06)$.
Since Algorithm~3 first passage times are greater than expected
network degree time Algorithm~4 first passage times, it must be that
a disproportionately large set of the secondary structures visited in
folding trajectories of the 10-mer GGGGGCCCCC have {\em larger} degree
(a larger number of neighboring structures that can be reached by the
addition or removal of a single base pair) than the network average.
This suggests that the (identical) folding trajectories of 
Algorithms~\ref{algo:3} and \ref{algo:4}
do {\em not} visit mainly low energy structures, although our detailed
analysis of one 10 nt RNA sequence can hardly be representative.

Note as well that the paired
times $T_c$ and $\langle N \rangle \cdot T_d$ are poorly correlated.
Since first passage times are not normally distributed, we compute 
the Spearman correlation of $0.439397$; although Pearson correlation should
not be used for non-normal distributions, the Pearson correlation is
$0.445950$.
Analysis of the same data after removal of outliers (e.g. computing the
10\%-trimmed mean, etc.) even strengthen the conclusions just reached
(data not shown).
It follows that the analogue of Theorem~\ref{thm1} for 
trajectories that terminate when reaching the MFE
does not hold -- moreover, the correlation is low for the paired trajectory
times for synchronized Algorithms~\ref{algo:3} and \ref{algo:4}
to reach the MFE structure.
Since tremendous computational resources would be required to perform
a similar analysis for longer RNA sequences, we restrict our attention to
the correlation between the synchronized and non-synchronized first
passage times for 

Algorithms~\ref{algo:3} (continuous-time event-driven Monte Carlo) 
and \ref{algo:4} (Gillespie)
-- see Tables~\ref{table:evan1000corr} and
\ref{table:DividingByExpNumNborsDoesNotHelp}.

\subsubsection*{Expected degree with respect to the uniform distribution}

Our software \cite{cloteJCC2015} also computes that the {\em uniform}
expected number of neighbors $\langle N \rangle = 3.548387 \approx 3.55$
for the set of all 62 secondary structures of the 10-mer GGGGGCCCCC,
defined by equation~(\ref{eqn:networkDegree}) where the free energy $E(s)$ 
of every secondary structure is defined to be zero, or equivalently by setting
probability $P(s) = \frac{1}{Z} = \frac{1}{62}$, where $Z$ now denotes the the 
total number of secondary structures of GGGGGCCCCC. We now re-analyze the
100,000 first passage times of Algorithms~\ref{algo:3} and \ref{algo:4}.

Recall that the null hypothesis $H_0$ asserts that
trajectory time $T_c$ for Algorithm~3
to reach the MFE structure is equal to $\langle N \rangle$ multiplied by
the trajectory time $T_d$ for Algorithm~4
to reach the MFE structure, except that we now use uniform expected
number of neighbors;
i.e. $H_0$ is the assertion that $T_c = 3.548387 \cdot T_d$.

The p-value for $n=100,000$
paired values $(T_c, \langle N \rangle \cdot T_d)$ is $3.42447E-09$.
For $\alpha=0.05$, to compute the $1-\alpha = 95\%$ confidence interval for 
$T_c-\langle N \rangle \cdot T_d$, we determine the 95\%
confidence interval of
$(\overline{x}_{dif}-E, \overline{x}_{dif}+E) = (-478.01,-269.87)$.
The 95\% confidence interval for an independent run of 
Algorithms~\ref{algo:3} and \ref{algo:4}
to generate another set of 100,000 first passage times was
$(-389.30, -362.03)$.
It follows that the uniform probability analogue of Theorem~\ref{thm1} for 
trajectories that terminate when reaching the MFE
does not hold.

Since Algorithm~3 first passage times are now {\em less} than the
{\em uniform} expected number of neighbors multiplied by 
Algorithm~4 first passage times, it must be that a
a disproportionately large set of the secondary structures which are visited in
folding trajectories of the 10-mer GGGGGCCCCC have {\em lower} degree
(a lower number of neighboring structures that can be reached by the
addition or removal of a single base pair) than the network average.
This suggests that the (identical) folding trajectories of 
Algorithms~\ref{algo:3} and \ref{algo:4}
do {\em not} visit mainly high energy structures (having few base pairs,
hence potentially a larger number of neighbors), although our detailed
analysis of one 10 nt RNA sequence can hardly be representative. 

Finally, we note that the ratio of the mean first passage time
$\langle T_c \rangle$ [resp.  $\langle T_d \rangle$],
taken over $n=100,000$ runs of Algorithm~3 [resp. Algorithm~4] is
$\frac{\langle T_c \rangle}{\langle T_d \rangle} \approx 3.43$, 
a value that appears somewhat closer to  uniform expected network degree
of $3.548387$ than the Boltzmann expected network degree of $3.031162$.

\subsubsection*{Correlation analysis for 20-mers}

In the case of RNA secondary structure folding kinetics, 
it appears that Gillespie MFPT times expected network
degree is even less correlated with Monte Carlo MFPT 
than unaltered Gillespie MFPT.
Table~\ref{table:evan1000corr} shows the Pearson correlation between the 
averages, taken over 1000 runs, of the MFPT for each of 1000 20 nt RNAs, 
when computed by the Monte Carlo (MC) method
(Algorithm~3$^{\dag}$), synchronized (syn) [resp. not synchronized (ns)] 
with Gillespie's (G) method (Algorithm~4$^{\dag}$). Additionally, we
performed two repetitions of the unsyncronized runs, designated experiment
A and B in the table.  For these data, the MFPT computed by
Monte Carlo is highly correlated with that computed by Gillespie:
G (syn) has correlation of 0.85702 with MC (syn) and a correlation of 
0.81030 with MC (ns).  Surprisingly, G (syn) has a somewhat {\em lower} 
orrelation of 0.85381, for MC* (syn), obtained by dividing the
MC (syn) time for each sequence by its expected number of neighbors; 
similarly, G (syn) has a somewhat {\em lower} correlation of 0.80559 
for MC* (ns), obtained by dividing the MC (ns) time for each sequence 
by its expected number of neighbors.  In other words, despite the fact that
the ratio of Monte Carlo trajectory time over Gillespie trajectory time
equals the expected number of neighbors for sufficiently long trajectories
as proved in Theorem~\ref{thm1}, 
there is no such relation between Monte Carlo MFPT
and Gillespie MFPT, presumably since the mean first passage time is 
generally reached within $K$ steps, where $K$ is too small for  the
asymptotic effects of Theorem~\ref{thm1} to apply.

\begin{table*}
\begin{tabular}{|l|llllll|}
\hline
&	{\mbox{\small MC (syn)}}	&{\mbox{\small MC (ns A)}}	&{\mbox{\small MC (ns B)}}	&{\mbox{\small G (syn)}}	&{\mbox{\small G (ns A)}}	&{\mbox{\small G (ns B)}}\\
\hline
{\mbox{\small MC (syn)}}	&1.000000	&0.994405	&0.981810	&0.857019	&0.871860	&0.880024\\
{\mbox{\small MC (ns A)}}	&0.994405	&1.000000	&0.989948	&0.810297	&0.830360	&0.839469\\
{\mbox{\small MC (ns B)}}	&0.981810	&0.989948	&1.000000	&0.783331	&0.809399	&0.814758\\
{\mbox{\small G (syn)}}	&0.857019	&0.810297	&0.783331	&1.000000	&0.996953	&0.996666\\
{\mbox{\small G (ns A)}}	&0.871860	&0.830360	&0.809399	&0.996953	&1.000000	&0.998148\\
{\mbox{\small G (ns B)}}	&0.880024	&0.839469	&0.814758	&0.996666	&0.998148	&1.000000\\
\hline
\end{tabular}
\caption{Correlation shown by folding time averages, 
taken over 1000 runs, for 1000
20 nt random sequences taken from the benchmarking set of 
\cite{Senter.jcb15}. Abbreviations are as follows.
MC (sym): synchronized Monte Carlo Algorithm~3$^{\dag}$;
MC (ns A): batch A for nonsynchronized Monte Carlo Algorithm~3$^{\dag}$;
MC (ns B): batch B for nonsynchronized Monte Carlo Algorithm~3$^{\dag}$;
G (sym): synchronized Gillespie Algorithm~4$^{\dag}$;
G (ns A): batch A for nonsynchronized Gillespie Algorithm~4$^{\dag}$;
G (ns B): batch B for nonsynchronized Gillespie Algorithm~4$^{\dag}$.
}
\label{table:evan1000corr}
\end{table*}

\begin{table*}
\begin{tabular}{|l|llllll|}
\hline
&	{\mbox{\small MC(syn)}}	&{\mbox{\small MC(ns)}}	&{\mbox{\small G(syn)}}	&{\mbox{\small G(ns)}}	&{\mbox{\small MC*(syn)}}	&{\mbox{\small MC*(ns)}}\\
\hline
{\mbox{\small MC(syn)}}	&1.000000	&0.994405	&0.857019	&0.871860	&0.985542	&0.978662\\
{\mbox{\small MC(ns)}}	&0.994405	&1.000000	&0.810297	&0.830360	&0.981410	&0.985471\\
{\mbox{\small G(syn)}}	&0.857019	&0.810297	&1.000000	&0.996953	&0.853810	&0.805594\\
{\mbox{\small G(ns)}}	&0.871860	&0.830360	&0.996953	&1.000000	&0.871133	&0.827705\\
{\mbox{\small MC*(syn)}}	&0.985542	&0.981410	&0.853810	&0.871133	&1.000000	&0.994563\\
{\mbox{\small MC*(ns)}}	&0.978662	&0.985471	&0.805594	&0.827705	&0.994563	&1.000000\\
\hline
\end{tabular}
\caption{Correlation shown by folding time averages, 
taken over 1000 runs, for 1000
20 nt random sequences taken from the benchmarking set of 
\cite{Senter.jcb15}. Abbreviations are as in Table~\ref{table:evan1000corr},
with the addition that MC* (syn) [resp. MC* (ns)] is produced by
dividing the Monte Carlo (Algorithm~3$^{\dag}$) folding time by the expected
sequence connectivity for synchronized [resp. nonsynchronized] computations.
}
\label{table:DividingByExpNumNborsDoesNotHelp}
\end{table*}

\section{Discussion}
\label{section:discussion}

In this paper, we have compared Markov chains and related Markov processes
by considering four closely related Algorithms 1,2,3,4.
If the underlying graph for the Markov chain $\mathbb{M}$ is $N$-regular, 
so that each state of $\mathbb{M}$ has exactly $N$ neighbors, then 
it follows that on average, mean first passage time (MFPT) 
computed by each of the Monte Carlo Algorithms 1,2 and 3  equals $N$ 
multiplied by the MFTP computed by Gillespie Algorithm 4.  Although the
Markov chain of RNA secondary structures of a given RNA sequence is 
generally not
$N$-regular for any $N$, the total time along a Monte Carlo trajectory
is asymptotically equal to the Boltzmann expected number of neighbors
$\langle N \rangle = \sum_{s} \frac{\exp(-E(s)/RT)}{Z} \cdot N(s)$ 
multiplied by 
the total time along a Gillespie trajectory, as proved in Theorem~\ref{thm1}.
Computational experiments on several RNAs confirm this result,
provided that trajectories are sufficiently long to exhibit
asymptotic properties of Markov chains. Our code {\tt mc.c}
for Algorithms 3 and 4, is written in C and makes calls to the
function {\tt energy\_of\_structure()} from {\tt libRNA.a} of
Vienna RNA Package \cite{Lorenz.amb11}. This program, along with
C programs to compute the expected network degree for RNA secondary
structures with move set $MS_1$ (base pair addition or deletion)
\cite{cloteJCC2015} or with move set $MS_2$ (base pair addition, deletion
or shift) \cite{Clote.po15}, is publicly available at
\url{http://bioinformatics.bc.edu/clotelab/RNAexpNumNbors}.

Since Anfinsen's pioneering experimental result on the folding of bovine
ribonuclease \cite{anfinsen}, it is widely accepted that the
native state of a biomolecule is a free energy minimum.
In the literature on biomolecular folding, there is sometimes 
a tacit assumption that kinetics simulations
using the Monte Carlo and Gillespie algorithms yield comparable results.
Indeed, in \cite{amatoRecombRNA} Tang et al. assert that
``We demonstrate with two different RNA that the different analysis methods
(ME, MC, MMC) produce comparable results and can be used 
interchangeably.\footnote{ME stands for the {\em Master Equation}, meaning the
computation of the time-dependent population occupancy vector 
$\mathbbp(t)$ by solution of the master equation 
$\frac{d \mathbbp(t)}{dt} = \mathbbp(t) \cdot Q$,
where $Q$ is the rate matrix. MC stands for Monte Carlo simulation, and 
MMC stands for {\em Map-based Monte Carlo} simulation, a method inspired
by probabilistic road map methods from robotics, where a constant $k$
many closest neighboring secondary structures are selected for each 
RNA secondary structure from a sampled collection of modest size.
Subsequently Monte Carlo simulation is applied to the resulting network,
which is much smaller than the network of all secondary structures.
The correctness of the authors' assertion is due uniquely to the fact 
that the roadmap sampling network is {\em $k$-regular}.}''
The results of this paper suggest that one should not make tacit
assumptions concerning folding kinetics simulations using Monte Carlo and
Gillespie algorithms, when different nodes in the network have different 
numbers of neighbors, as in the case for RNA secondary structures.

If computing the mean first passage time requires sufficiently long 
trajectories, then we would expect Monte Carlo MFPT to approximately equal
Gillespie MFPT times the expected number of neighbors.  However, for a
slight modification of Example~\ref{example1} 
described in Example~\ref{example2}, as well as for
a benchmarking set of 1000 20 nt RNAs, each of which has at most 2,500
secondary structures, no such relation was found.  Even worse, for the
set of 1000 RNAs, the
correlation between Monte Carlo MFPT and Gillespie MFTP times
the expected number of neighbors was found to be {\em lower} 
than the correlation without its consideration.  This result suggests
that number of trajectory 
steps necessary to reach the minimum free energy
structure may be too small to see the asymptotic relation expected by
Theorem~\ref{thm1}. We conclude that RNA secondary structure folding 
may occur {\em faster} than the time scale required for Theorem~\ref{thm1}, 
a type of {\em Levinthal paradox} \cite{levinthal:1968} which suggests that
{\em folding pathways} could be encoded in RNA sequences, or that
RNA secondary structure formation could follow either a 
{\em kinetic funnel model} \cite{Bryngelson.p95} or
a {\em kinetic hub model} \cite{Bowman.pnas10}.

Next, we argue that macromolecular folding kinetics are better captured
by Gillespie's Algorithm 4, than Monte Carlo Algorithms 1,2,3.
For all but pathological
RNA sequences, it is the case that detailed balance 
$\pi_x \cdot p_{x,y} = \pi_y \cdot p_{y,x}$ does not hold for 
the Boltzmann distribution $\pi$,
as shown by considering $x$ to be the empty structure and $y$ to be a structure
containing exactly one base pair. Since the Boltzmann distribution is
not necessarily the stationary distribution, this argument does not 
imply that detailed balance does not hold; however in 
Section~\ref{section:foldingTrajectoriesFor10mer}
the (non-Boltzmannian) stationary distribution is computed for a
tiny RNA sequence, for which it is shown that detailed balance does not
hold.

Since the stationary distribution for the Markov chain underlying
each of the Monte Carlo Algorithms 1,2,3 is
not necessarily the Boltzmann distribution, one might instead 
consider modified versions of these algorithms defined as follows.
Algorithm 1$^{\ddag}$ is obtained from Algorithm 1$^{\dag}$ by replacing
the expression in line 9 by
\[
\frac{\pi_y}{\pi_x} \cdot \frac{N_x}{N_y}
\]
or equivalently by
\[
\exp\left( \frac{-(E(y)-E(x))}{RT} \right) \cdot \frac{N_x}{N_y}.
\]
Similarly,
Algorithm 2$^{\ddag}$ [resp. 3$^{\ddag}$] 
is obtained from Algorithm 2$^{\dag}$ [resp. 3$^{\dag}$]
by replacing the expression in line 9 by
\[
\frac{1}{N_x} \cdot \min\big(1, \frac{\pi_y}{\pi_x} \cdot \frac{N_x}{N_y} 
\big)
\]
or equivalently
\[
\frac{1}{N_x} \cdot \min\big(1, 
\exp\left( \frac{-(E(y)-E(x))}{RT} \right) \cdot \frac{N_x}{N_y} \big).
\]
This ensures that the new Markov chain underlying each of the Algorithms
1$^{\ddag}$, 2$^{\ddag}$, 3$^{\ddag}$ is Markov chain $\mathbb{M}_H$,
whose transition probability matrix $P_H = (p^H_{x,y})$ is defined by
Hastings' trick, i.e.
\begin{align}
\label{eqn:hastingsTrick}
p^{H}_{x,y} &= \frac{1}{N_x} \min(1, \exp(-(E(y)-E(x))/RT) \cdot N_x/N_y)
\end{align}
However, the embedded matrix (jump matrix) for
$M_H$ would then be different for the Markov process $\mathbb{M}_2$
underlying the Gillespie algorithm. Although we would have ensured that
the stationary distribution for each of the Algorithms
1$^{\ddag}$, 2$^{\ddag}$, 3$^{\ddag}$, and 4 is the Boltzmann distribution,
we would no longer have trajectory times related as in Theorem~\ref{thm1}.

By Theorem~1, either the Monte Carlo algorithm or the Gillespie algorithm
may be used for statistical analysis of trajectories (frequency of visitation
of states, etc.). However in the context of RNA secondary structure folding
time to the minimum free energy structure, the Boltzmann distribution 
is the stationary distribution {\em only} for the Gillespie Algorithm 4 
and for the Monte Carlo Algorithms 2,3 modified by Hastings'
trick using transition probabilities from equation~(\ref{eqn:hastingsTrick}).
Moreover, the minimum free energy structure is {\em targeted} only if the 
Boltzmann distribution is the stationary distribution, where by {\em targeted},
we mean that the MFE structure has the highest probability, and that
$\lim\limits_{n \rightarrow \infty} Pr[X_n=x] = \pi_x$, where $X_n$ is the
random variable for the secondary structure at step $n$ and $\pi_x$ is
the Boltzmann probability of structure $x$. However, there is a
serious computational cost for the Hastings-modified Monte Carlo algorithm;
indeed, determination of the update to state $x$ requires the computation
of free energies of all neighbors of not only current state $x$, but
as well of all neighbors in $N_y$ of all states $y \in N_x$. For repeated
RNA kinetics simulations, this cost could be prohibitive. 

We argue that the Gillespie algorithm should be used for  RNA secondary
structure folding kinetics, and that specifically one should compute 
the population occupancy vector $\mathbbp(t)$ determined by solution of the
matrix differential equation $\frac{d \mathbbp(t)}{dt} = \mathbbp(t)
\cdot  Q$, for rate matrix $Q$ as in \cite{Zhang.pnas02,Xu.pnas16},
rather than the mean first passage time by use of the fundamental matrix
or by matrix inversion \cite{meyerMFPT} as done for
Markov state models in \cite{Bowman.pnas10,Huang.psb10}. For
synthetic design of RNA molecules \cite{Zadeh.jcc11,Dotu.nar15},
we advocate fast MFPT computation using coarse-grained models 
\cite{wolfingerStadler:kinetics,Senter.jcb15} to select design candidates 
for subsequent scrutiny by more accurate methods such as {\tt KFOLD}.

\section{Acknowledgements}

This research was begun during a visit to California Institute of
Technology, Free University of Berlin and the Max Planck Institute
for Molecular Genetics.  For discussions, P.C. would like 
to thank Frank No{\'e}, Knut Reinert, Martin Vingron
and Marcus Weber (Berlin) and Niles Pierce, Erik Winfree  (Pasadena).
The research was funded by a Guggenheim Fellowship, Deutscher Akademischer
Austauschdienst (DAAD), and by National Science Foundation grant DBI-1262439.
Any opinions, findings,
and conclusions or recommendations expressed in this material are
those of the authors and do not necessarily reflect the views of the
National Science Foundation.

\hfill\break \clearpage
\begin{figure*}
\begin{algorithm}[\bf Discrete-time time-driven MC algorithm for MFPT]
\label{algo:1}
\hfill\break
\medskip

\begin{small}
\mverbatim
 1. procedure Metropolis(initialState $x_0$, targetState $x_{\infty}$)
 2.   $x = x_0$; time $t=0$
 3.   while $x \ne x_{\infty}$ \{
 4.     choose random $y \in N_x$ 
 5.     if $E(y)<E(x)$  //greedy step
 6.       x = y          //update x
 7.     else               //Metropolis step
 8.       choose random $z \in (0,1)$
 9.       if $z < \exp\left(-\frac{E(y)-E(x)}{RT}\right)$
10.         x = y      //update x
11.     t = t + 1  //update time regardless of whether x is modified
12.    \} //end of while loop
13.   return final state $x$, total time $t$
|mendverbatim
\end{small}
\end{algorithm}
\caption{Discrete-time time-driven Monte Carlo (MC) algorithm to 
estimate mean first passage time for a given Markov chain, whose
transition probabilities are given by equation~(\ref{eqn:transitionProb1}).
Since the Boltzmann probabilities are defined by $p_x^*=\exp(-E(x)/RT)/Z$,
$p_y^*=\exp(-E(y)/RT)/Z$, it follows that $p_y^*/p_x^* = \exp(-(E(y)-E(x))/RT)$,
so that the probability that state $x$ is modified to $y$ in
lines 5-10 is precisely $p_{x,y} = 1/N_x \cdot \min(1,\exp(-(E(y)-E(x))/RT))$.
}
\label{fig:algo1}
\end{figure*}

\begin{figure*}
\begin{algorithm}[\bf Discrete-time event-driven MC algorithm for MFPT]
\label{algo:2}
\hfill\break
\medskip

\begin{small}
\mverbatim
 1. procedure Metropolis(initialState $x_0$, targetState $x_{\infty}$)
 2.   $x=x_0$; time $t=0$
 3.   while $x \ne x_{\infty}$ \{
 4.     $\Phi_1=0$; //probability of leaving current state $x$
 5.     for all $y \in N_x$ do
 6.       $p_{x,y} = \frac{1}{N_x} \min\left(1,\exp(-\frac{E(y)-E(x)}{RT})\right)$ 
 7.       $\Phi_1 += p_{x,y}$
 8.     for all $y \in N_x$ do $s_{x,y} = \frac{p_{x,y}}{\Phi_1}$ //jump probabilities
 9.     sample $\Delta t \thicksim Geom(\frac{1}{\Phi},\sqrt{\frac{1-\Phi}{\Phi^2}})$ where $\Phi=\Phi_1$
10.     t += $\Delta t$ //update time
11.     choose random $z_2 \in (0,1)$
12.     //sample $y\in N_x$, using roulette wheel as in the following
13.     sum = 0
14.     for $y \in N_x$
15.       sum = sum + $s_{x,y}$
16.       if $z_2 \leq sum$ then
17.         x=y; break //use roulette wheel
18.    \}  //end of while loop
19.   return final state $x$, total time $t$
|mendverbatim
\end{small}
\end{algorithm}
\caption{Discrete-time event-driven Monte Carlo (MC) algorithm to 
estimate mean first passage time for a given Markov chain, whose
transition probabilities are given by equation~(\ref{eqn:transitionProb1}).
Since the probability that state $x$ is modified to $y$ in
lines 5-10 of Algorithm~\ref{algo:1} equals
$p_{x,y} = 1/N_x \cdot \min(1,\exp(-(E(y)-E(x))/RT))$, it follows that
the probability that $x$ will be modified is $\sum_{z \in N_x} p_{x,z}$
which equals $1-p_{x,x}$. The probability that $x$ will be modified in
the $k$th iteration of the while loop of Algorithm~\ref{algo:1}
(but not before) is exactly $p_{x,x}^{k-1} \cdot (1-p_{x,x})$; i.e.
equivalent to the probability that the heads is obtained in the
$k$th coin flip (but not before) for a coin whose heads probability is
$(1-p_{x,x})$, given as $\Phi$ in line 7 of Algorithm~\ref{algo:2}. 
It follows that the number of steps $\Delta t$ before
the state $x$ is modified is geometrically distributed with parameter
$\Phi=(1-p_{x,x})$; since the mean [resp. standard deviation] for a 
geometrically distributed random variable with parameter $\Phi$ is
$\frac{1}{\Phi}$ [resp.  $\sqrt{\frac{1-\Phi}{\Phi^2}}$] we denote the
distribution by $Geom(\frac{1}{\Phi},\sqrt{\frac{1-\Phi}{\Phi^2}})$. 
Time increments $\Delta t$ are sampled from this geometric distribution,
hence the expected time increment in line 9 of Algorithm~\ref{algo:2}
is identical to the expected time increment before state $x$ is modified
in lines $4-10$ of Algorithm~\ref{algo:1}, so both algorithms are
equivalent.
}
\label{fig:algo2}
\end{figure*}

\begin{figure*}
\begin{algorithm}[\bf Continuous-time event-driven MC algorithm for MFPT]
\label{algo:3}
\hfill\break
\medskip

\begin{small}
\mverbatim
 1. procedure hybridMC(initialState $x_0$, targetState $x_{\infty}$)
 2.   $x=x_0$; time $t=0$
 3.   while $x \ne x_{\infty}$ \{
 4.     $\Phi_1=0$ //$\Phi_1$ is probability of leaving $x$
 5.     for all $y \in N_x$ do
 6.       $p_{x,y} = \frac{1}{N_x} \min\left(1,\exp(-\frac{E(y)-E(x)}{RT})\right)$
 7.       $\Phi_1 += p_{x,y}$
 8.     for all $y \in N_x$ do $s_{x,y} = \frac{p_{x,y}}{\Phi_1}$ //jump probabilities
 9.     //sample $\Delta t \thicksim Exp(\frac{1}{\Phi},\frac{1}{\Phi})$ where $\Phi=\Phi_1$
10.     choose random $z_1 \in (0,1)$
11.     $\Delta t$ = $-\frac{1}{\Phi_1} \ln(z_1)$    //sample time increment
12.     t += $\Delta t$                            //update time
13.     choose random $z_2 \in (0,1)$
14.     sum = 0
15.     for $y \in N_x$
16.       sum += $s_{x,y}$
17.       if $z_2 \leq sum$ then
18.         x=y; break //use roulette wheel
19.    \} //end of while loop
20.   return final state $x$, total time $t$
|mendverbatim
\end{small}
\end{algorithm}
\caption{Continuous-time event-driven Monte Carlo (MC) algorithm, to
estimate mean first passage time for a given Markov chain, whose
transition probabilities are given by equation~(\ref{eqn:transitionProb1}).
The geometically distibuted time increment $\Delta t$ from line 9 of
Algorithm~\ref{algo:2} is replaced in line 9 of the current algorithm
by the exponentially distributed time increment having the same expectation 
$\frac{1}{\Phi}$. As the standard deviatioin of the exponential distribution
with parameter $1/\Phi$ is also $1/\Phi$, we denote this distribution by
$Exp(\frac{1}{\Phi},\frac{1}{\Phi})$ -- this allows the reader to immediately
see that the time increment $\Delta t$ in line 9 of both algorithms has
the same mean and approximately the same standard deviation (especially if
$\Phi$ is small).
}
\label{fig:algo3}
\end{figure*}

\begin{figure*}
\begin{algorithm}[\bf Gillespie's algorithm for MFPT]
\label{algo:4}
\hfill\break
\medskip

\begin{small}
\mverbatim
 1. procedure Gillespie(initialState $x_0$, targetState $x_{\infty}$)
 2.   $x = x_0$; time $t=0$
 3.   while $x \ne x_{\infty}$ \{
 4.     $\Phi_2=0$ //$\Phi_2$ is flux (not probability) out of $x$
 5.     for all $y \in N_x$ do
 6.       $q_{x,y} = \min\left(1,\exp(-\frac{E(y)-E(x)}{RT})\right)$
 7.       $\Phi_2 += q_{x,y}$
 8.     for all $y \in N_x$ do $s_{x,y} = \frac{q_{x,y}}{\Phi_2}$ //jump probabilities
 9.     //sample $\Delta t \thicksim Exp(\frac{1}{\Phi},\frac{1}{\Phi})$ where $\Phi=\Phi_2$
10.     choose random $z_1 \in (0,1)$
11.     $\Delta$ t = $-\frac{1}{\Phi_2} \ln(z_1)$  //sample time increment
12.     t += $\Delta$ t                          //update time
13.     choose random $z_2 \in (0,1)$
14.     sum = 0
15.     for $y \in N_x$
16.       sum += $s_{x,y}$
17.       if $z_2 \leq sum$ then
18.         x=y; break //use roulette wheel
19.    \} //end of while loop
20.   return final state $x$, total time $t$
|mendverbatim
\end{small}
\end{algorithm}
\caption{Continuous-time event-driven Gillespie algorithm, to
estimate mean first passage time for a given Markov {\em process}, whose
transition {\em rates} are given by equation~(\ref{eqn:transitionProb2}).
The only difference between this algorithm and Algorithm~\ref{algo:3}
is that in line 6, the transition rate $q_{x,y}$ is missing the 
factor $1/N_x$ present in transition probability $p_{x,y}$, and so
the value $\Phi$ from line 9 is smaller in the current algorithm than 
in Algorithm~\ref{algo:3} by the factor $N_x$. It immediately follows that
Algorithms~\ref{algo:3} and \ref{algo:4} are identical in a regular
network where $N_x = N$ for all states $x$; moreover, it is suggestive
(but incorrect for MFPT) that the values $\Phi$ from line 9 of Algorithms
\ref{algo:3} and \ref{algo:4} might be related by the Boltzmann expected
number $\langle N_x \rangle$ of neighbors.
}
\label{fig:algo4}
\end{figure*}

\begin{figure*}
\includegraphics[width=0.35\linewidth]{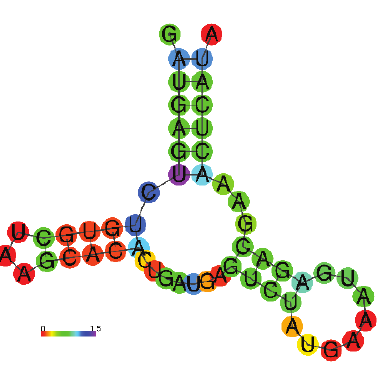}
\hspace{1cm}
\includegraphics[width=0.65\linewidth]{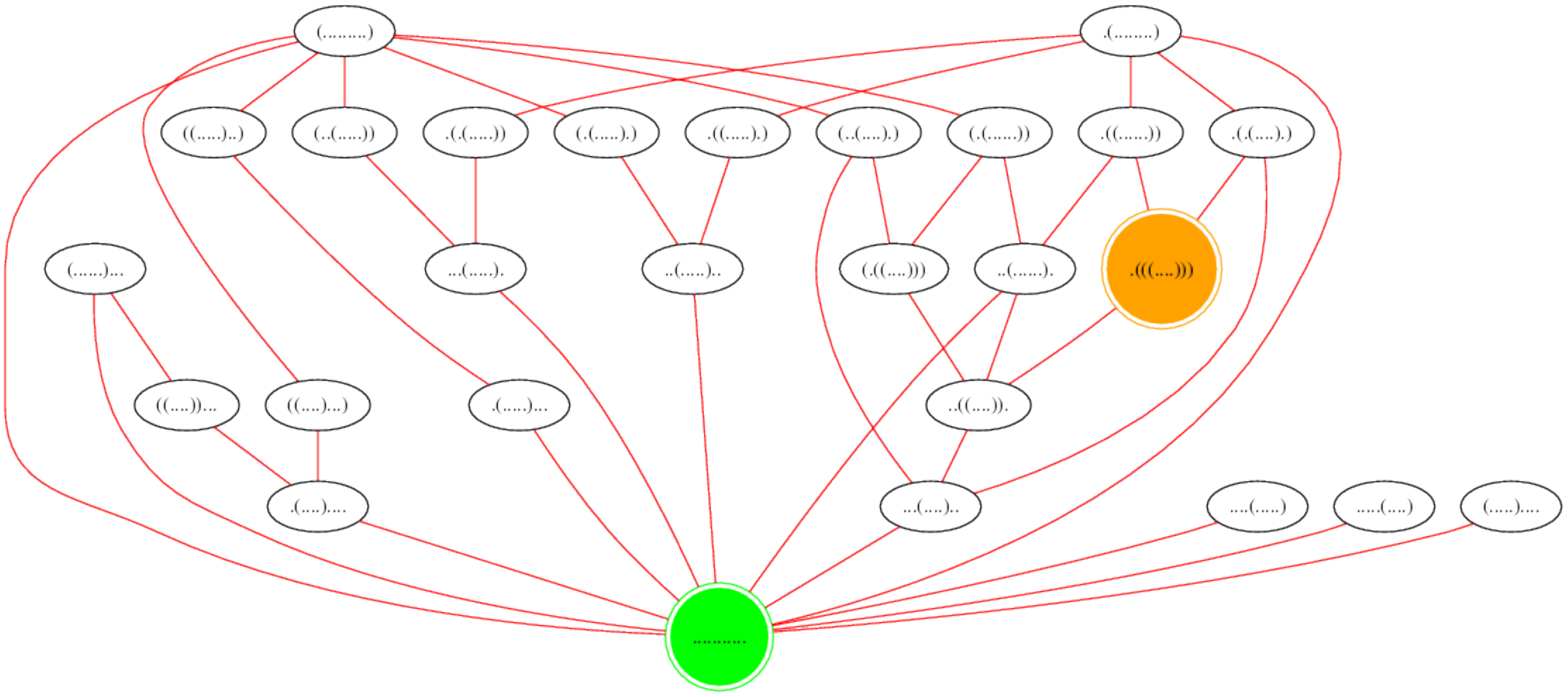}
\caption{{\em (Left)} Minimum free energy secondary (MFE) structure of
the 54 nt Peach Latent Mosaic Viroid (PLMVd) AJ005312.1/282-335,
colored by positional entropy -- see \cite{Huynen.jmb97} for the definition
of positional entropy. The MFE structure was computed by {\tt RNAfold} using
Vienna RNA Package 1.8.5, employing the Turner 1999 free energy
parameters. This structure is identical to the consensus structure from
the Rfam database \cite{Gardner.nar11}, while the structure produced by
Vienna RNA Package 2.1.9, employing the Turner 2004 free energy
parameters consists of two external stem-loops.  This example shows the 
sometimes dramatic difference in
MFE structure computation due solely to slightly different
free energy parameters  \cite{Turner.nar10}. 
{\em (Right)} Network of secondary structures for the 11-nt RNA sequence
{\tt GGCCGGCCGGC}, where edges are indicated between every two structures
whose base pair distance is $1$ (i.e. move set $MS_1$). 
The expected degree for this network having 27 nodes and 42 edges is
$3.111111$ with respect to the uniform distribution
and $3.074169$ with respect to the Boltzmann distribution.
The MFE structure of {\tt GCGCGCGCGCG}, depicted in the orange circle,
has free energy $-2.6$ kcal/mol. Mean first passage time for secondary
structure folding is the average first passage time from the empty initial
structure (green circle) to the MFE structure (orange circle), where
edge probabilities [rates] for the Monte Carlo [resp. Gillespie] algorithm
is given by equation~(\ref{eqn:transitionProb1})
[resp. equation~(\ref{eqn:transitionProb2})].
}
\label{fig:Vienna185andVienna207}
\end{figure*}

\begin{figure*}[!ht]
\begin{center}
\includegraphics[width=.45\linewidth]{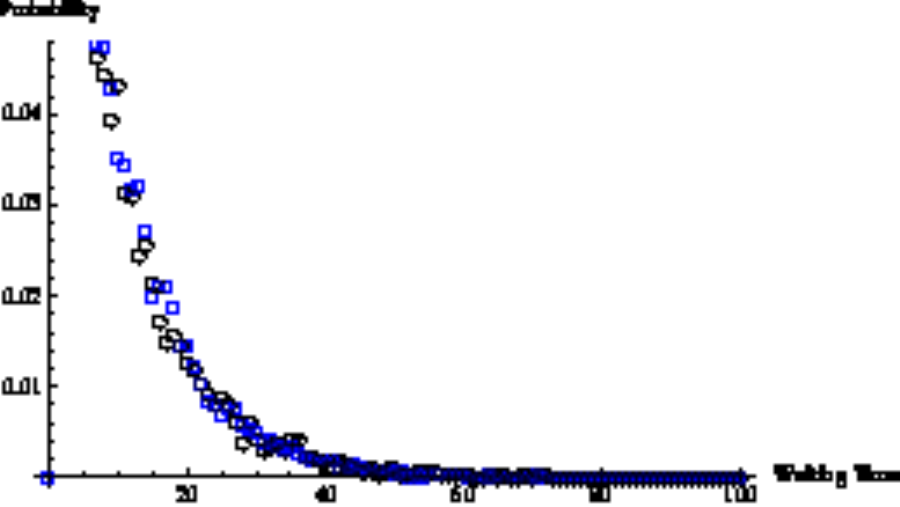}
\includegraphics[width=.45\linewidth]{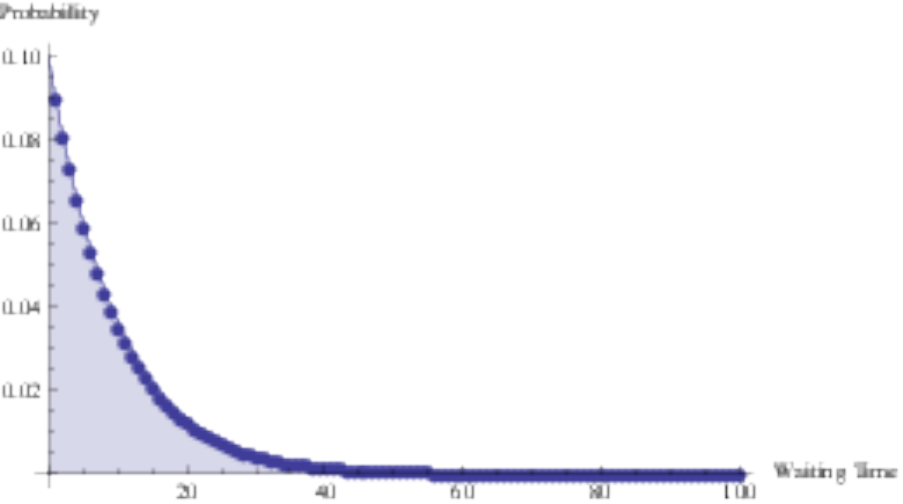}
\caption{{\em (Left)} Superimposition of histograms for the geometric and
exponential distributions. In each case, using the method described in
the text, 10,000 points were sampled from the geometric (blue squares)
[resp. exponential (black circles)]
distribution having mean 10 and standard deviation $\sqrt{90} \approx
9.4868$ [resp. mean 10]. For the geometric distribution, this corresponds
to the waiting time before obtaining a heads, where the probability of
flipping a heads is $0.1$. For the 10,000 points sampled from the
geometric distribution, the sample mean
is 9.995400 and the sample standard deviation is 9.413309. For the 10,000
points sampled from the exponential distribution, the sample mean is
9.989983 and the sample standard deviation is 9.950308.  
{\em (Right)} Superimposition of the probability [resp. probability
density] of the geometric [resp. exponential] distribution having mean
of 10, as computed by Mathematica. 
Graphics produced with Mathematica.
}
\label{fig:waitingTimeHistogramGeomertricAndExponentialDistribution}
\end{center}
\end{figure*}

\begin{figure*}[!ht]
\begin{center}
\includegraphics[width=0.45\linewidth]{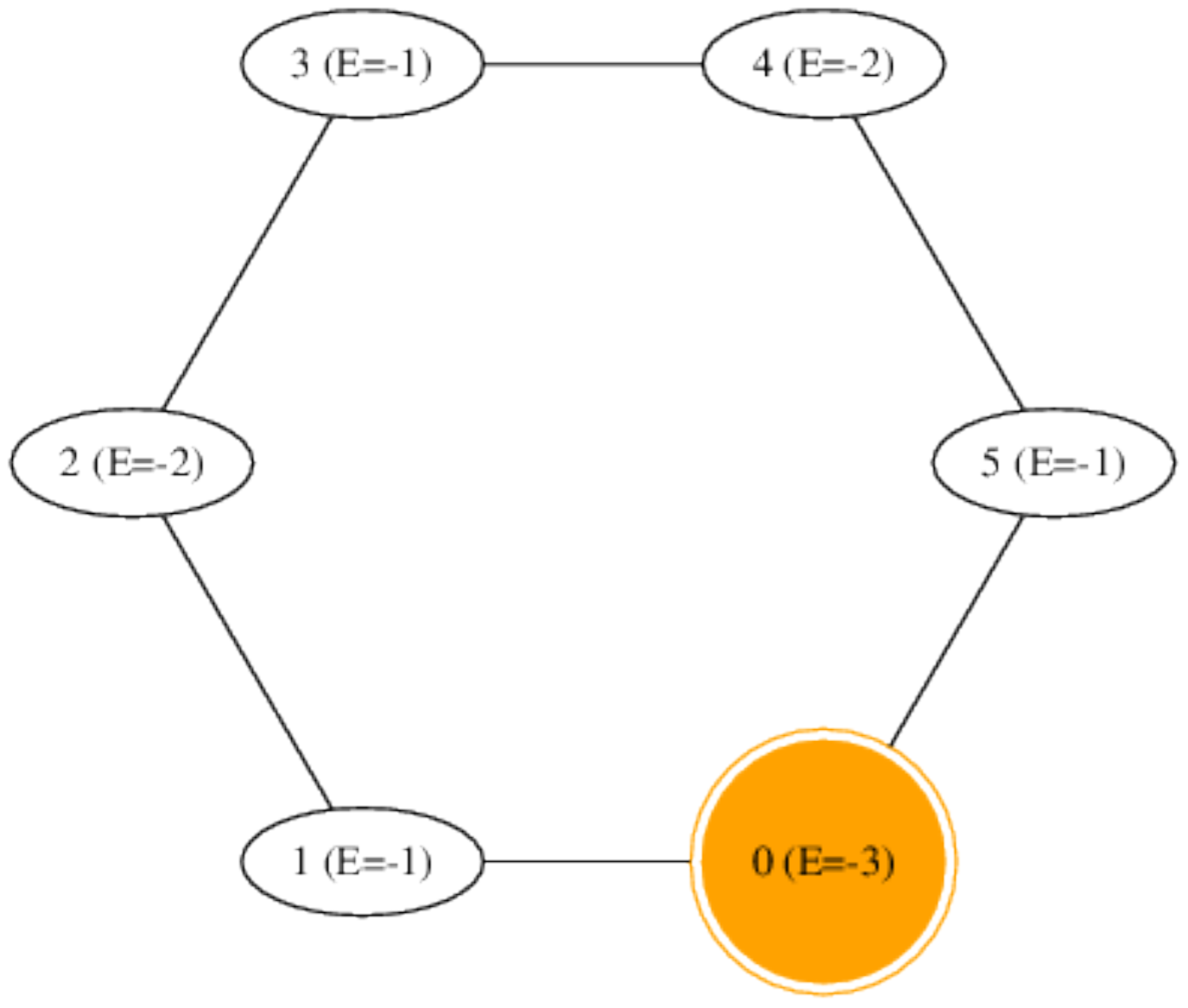}
\includegraphics[width=0.45\linewidth]{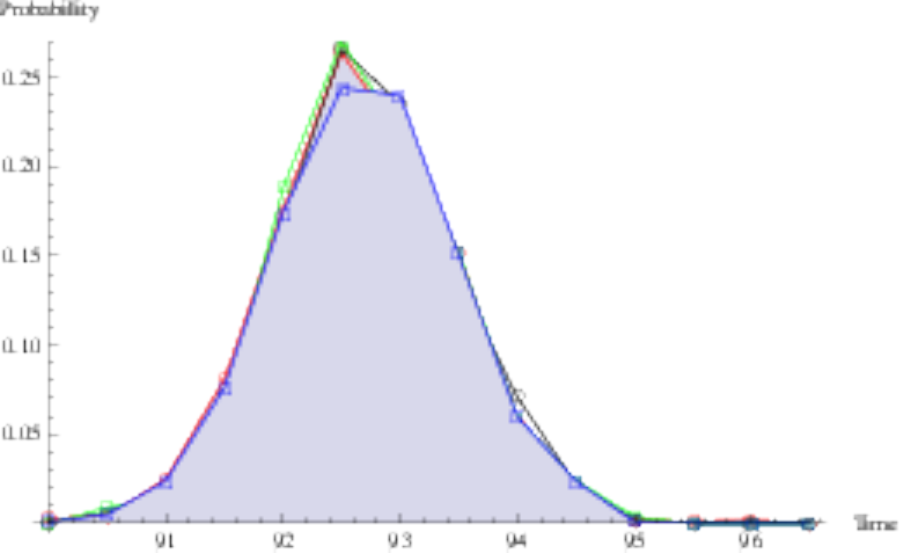}
\caption{
{\em (Left)} Topology of Markov chain/process from Example~\ref{example1},
consisting of $2n$ states $0,1,\ldots,2n-1$, 
each having exactly two neighbors. State $0$ has energy
$-3$, non-zero even [resp. odd] states have energy $-2$ [resp. $-1$].
The figure illustrates the topology for $n=3$, 
Example~\ref{example1} in the text considers $n=10$.
{\em (Right)}
Superimposed histograms for Monte Carlo and Gillespie first passage times
for Example~\ref{example1}, taken over 1000 computational experiments, 
each experiment consisting of taking the average 
{\em first passage time} in 10,000 repetitions of
each of the following algorithms:
(1)  time-driven Monte Carlo algorithm (blue squares with filled area),
(2) event-driven Monte Carlo algorithm with geometrically distributed waiting 
times (black circles),
(3) event-driven Monte Carlo algorithm with exponentially distributed waiting 
times (green squares),
(4) Gillespie's algorithm (red circles). Histograms for a single experiment
for each algorithm, consisting of 10,000 separate runs, resembled
an exponential distribution, of course, but the averages of each experiment
are approximately normally distributed (central limit theorem).
}
\label{fig:comparisonFourMCalgorithms}
\end{center}
\end{figure*}

\begin{figure*}[!ht]
\centering
\includegraphics[width=.45\linewidth]{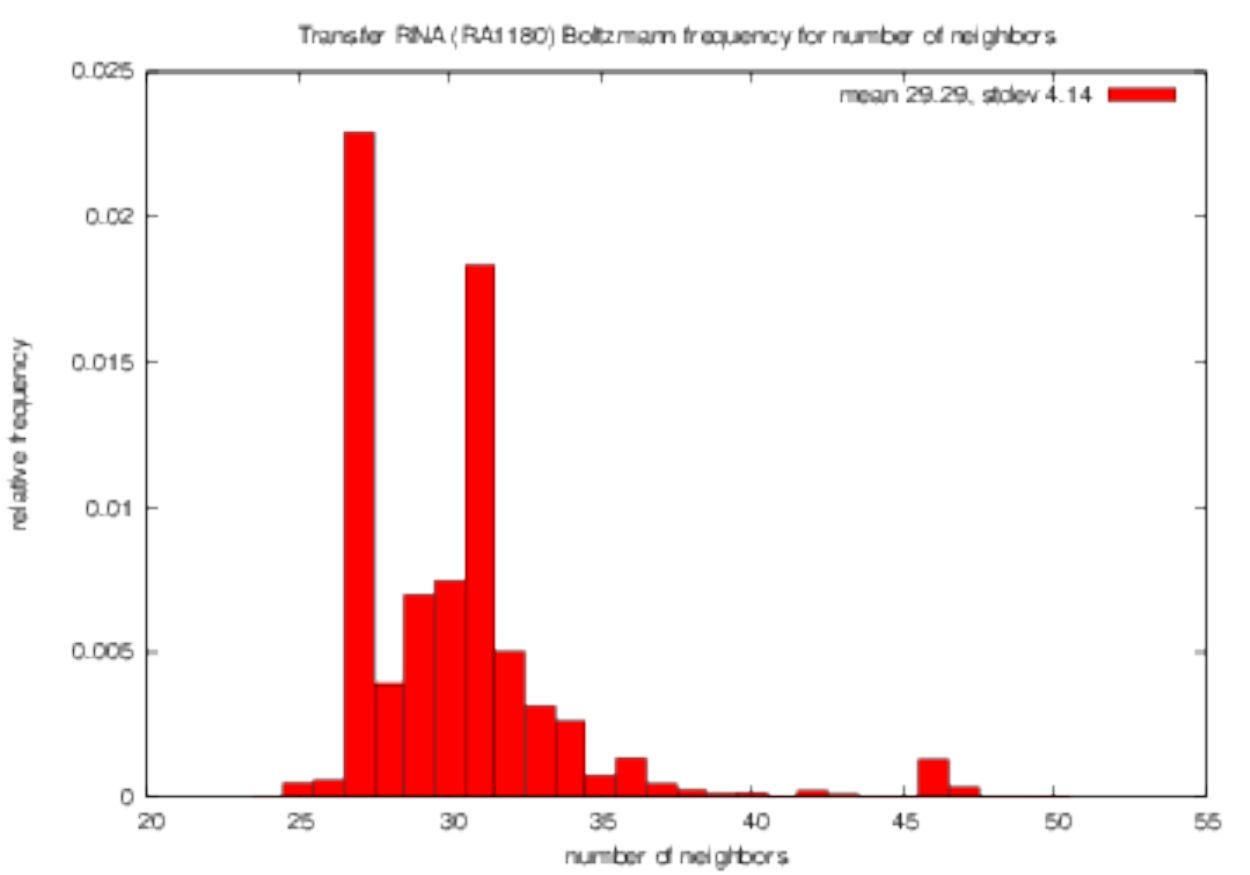}
\includegraphics[width=.45\linewidth]{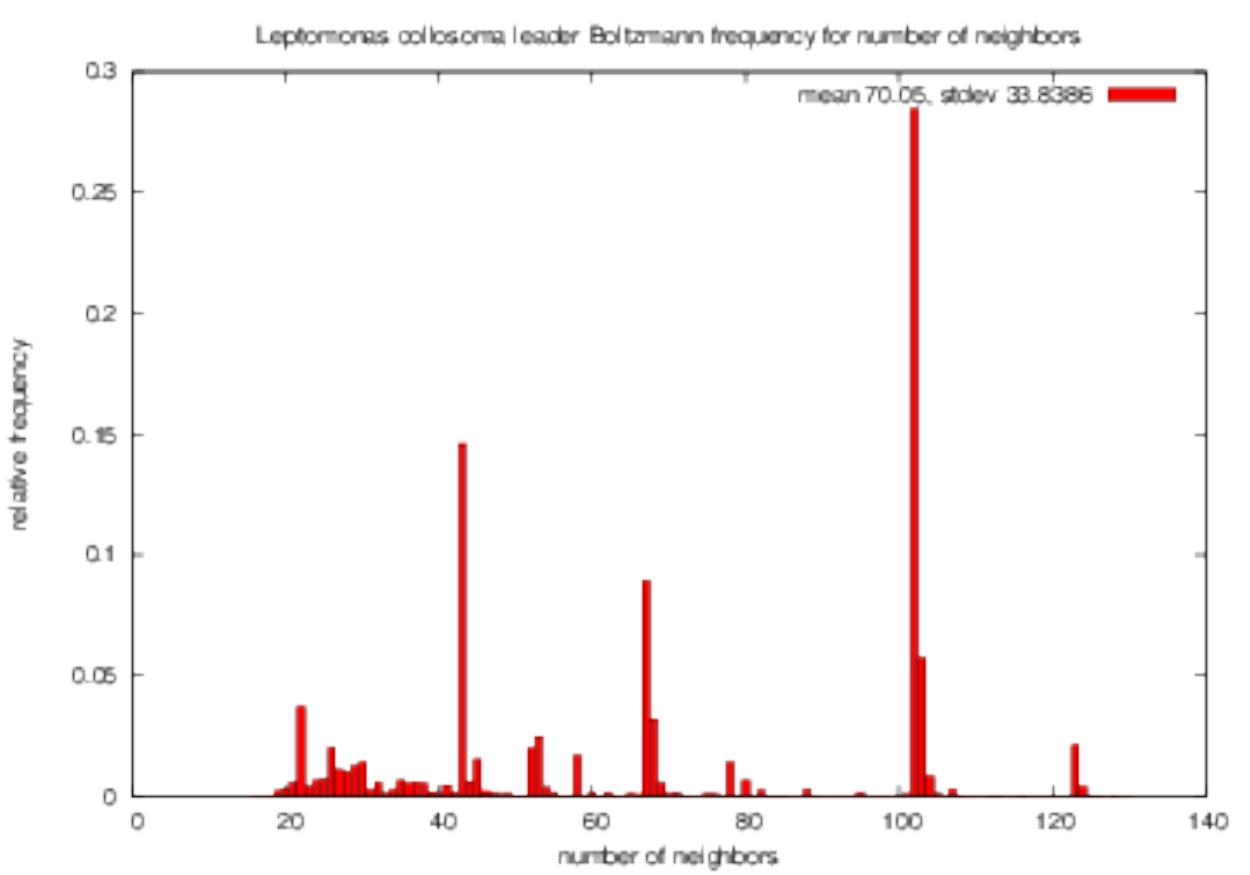}
\caption{Distribution of the Boltzmann-weighted 
number of {\em neighbors}, where the number of neighbors of
structure $s$ is the number of structures $t$ obtained from $s$ by
removal or addition of a single base pair.
Relative frequency at position $x$
is defined as the sum, taken over all {\em sampled}
structures $s$ of degree $x$, of the Boltzmann probability
$\exp(-E(s)/RT)/Z^*$,
where $Z^*$ is the sum of Boltzmann factors of all {\em sampled} structures.
This value approximates the true Boltzmann probability
$\exp(-E(s)/RT)/Z$, where the partition function $Z$ is the sum of
Boltzmann factors of {\em all} structures.
{\em (Left)} Distribution of the number of neighbors for transfer RNA 
(RA1180 from tRNAdb 2009 \cite{Juhling.nar09}), having expected number
of structures
$\langle N \rangle = 29.29 \pm 4.14$.  Using {\tt RNAsubopt -e 12}
\cite{Lorenz.amb11}, all structures were generated, whose free energy is within
12 kcal/mol of the minimum free energy ($110,572$ structures), and the
ratio $Z^*/Z$ of the sum of corresponding Boltzmann factors divided by
the partition function is $0.9990$.
{\em (Right)} Distribution of the number of neighbors
for spliced leader RNA, an RNA conformational switch, from the 
trypanosome {\em Leptomonas collosoma}, having expected number of 
structures $\langle N \rangle = 70.05 \pm 33.84$.
A total of 57,803 structures were generated using {\tt RNAsubopt -e 10}, 
and the ratio $Z^*/Z$ of the sum of corresponding Boltzmann factors divided by
the partition function is $0.9992$. 
}
\label{fig:RNAsuboptNumNbors}
\end{figure*}

\begin{figure*}[!ht]
\begin{center}
\includegraphics[width=0.45\linewidth]{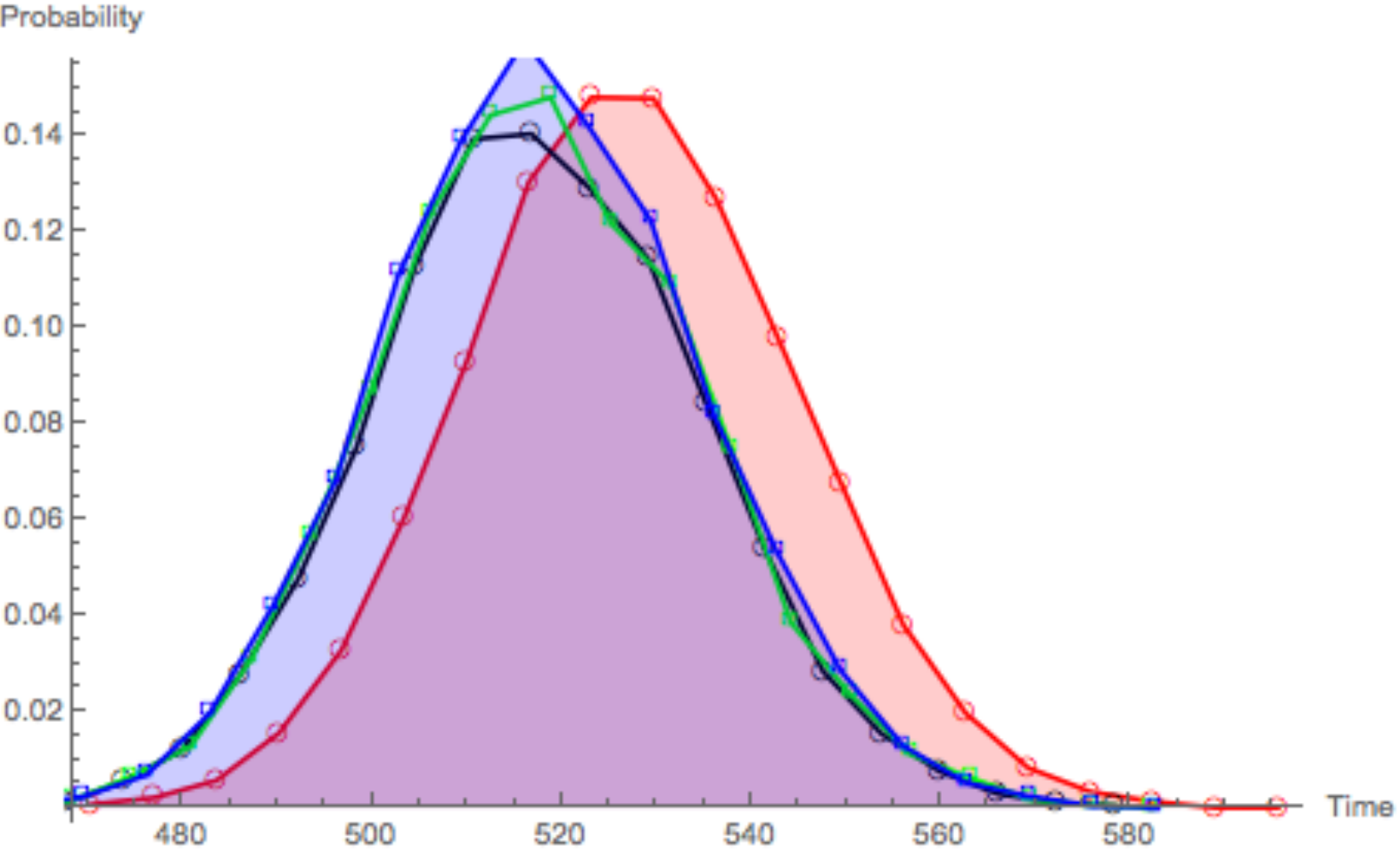}
\includegraphics[width=0.45\linewidth]{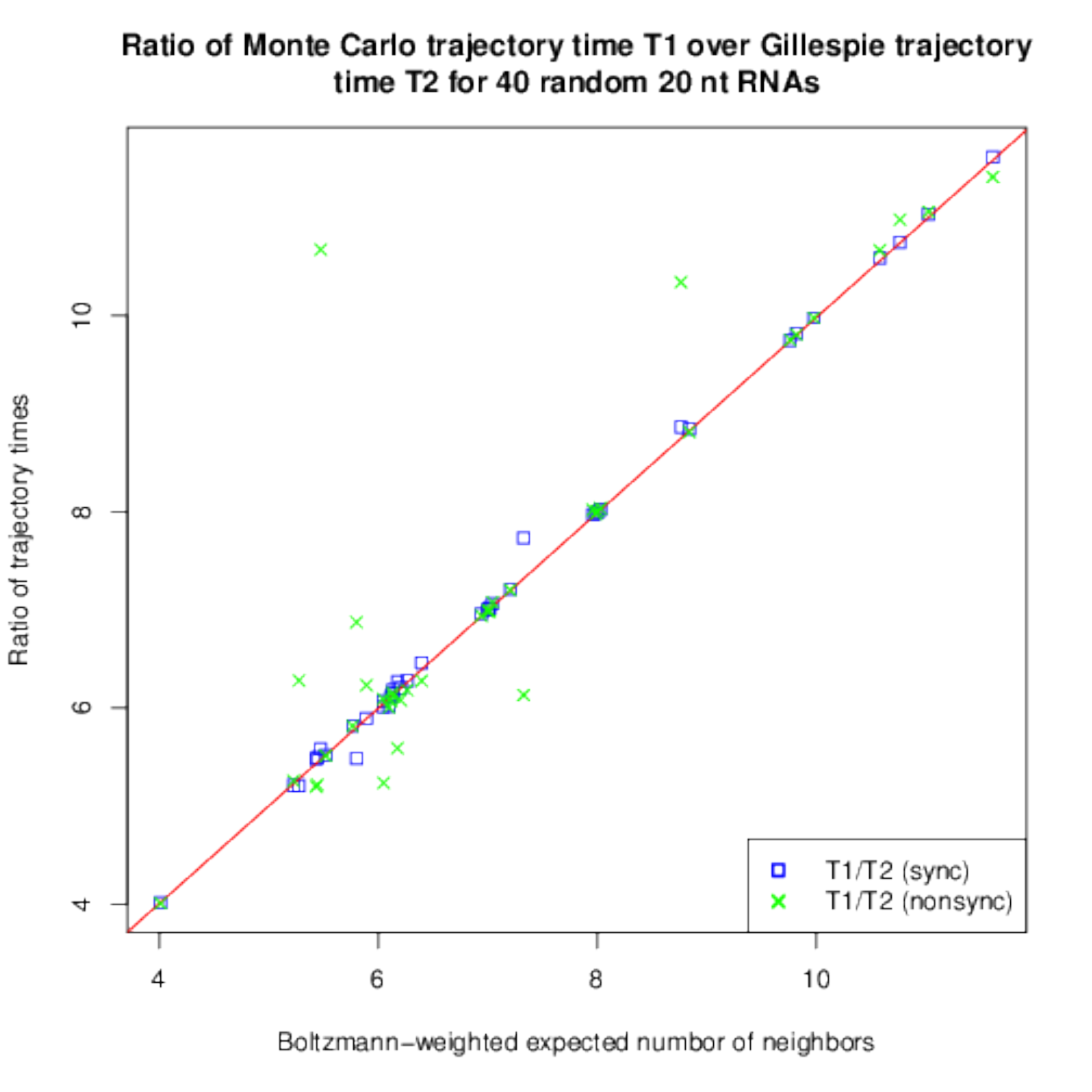}
\caption{
{\em (Left)}
Superimposed histograms for Monte Carlo and Gillespie first passage times
for Example~\ref{example2}, taken over 1000 computational experiments, 
each experiment consisting of taking the 
average {\em first passage time} in 10,000 repetitions of
each of the algorithms:
(1)  time-driven Monte Carlo algorithm (blue squares with filled area),
(2) event-driven Monte Carlo algorithm with geometrically distributed waiting 
times (black circles),
(3) event-driven Monte Carlo algorithm with exponentially distributed waiting 
times (green squares),
(4) Gillespie's algorithm (red circles). 
The computations were performed exactly as in 
Figure~\ref{fig:comparisonFourMCalgorithms} -- the topology of 
Example~\ref{example2} differs from that  of Example~\ref{example1}  only
in that there is no edge between the minimum energy state $0$ and the
last state $2n-1$, i.e. state 19 in Example~\ref{example2}.
{\em (Right)}
Ratio $\frac{\tau^K_1}{\tau^K_2}$ of trajectory time
for $K$ steps ($K=1$ million) in Monte Carlo and Gillespie algorithms,
plotted as a function of
the expected number of neighbors (expected degree) for 40 randomly generated
20 nt RNAs. This figure illustrates Theorem 1, which states that the
ratio $\frac{E(\tau^K_1)}{E(\tau^K_2)}$ of expected trajectory time for
$K$-step trajectories using Monte Carlo Algorithm 3$^{\dag}$ over the
expected trajectory time of $K$-step trajectories using Gillespie's Algorithm
4$^{\dag}$ equals the expected degree of the network of RNA secondary 
structures, as $K$ increases to infinity. Both synchronized and unsynchronized
trajectories are plotted.
}
\label{fig:comparisonFourMCalgorithmsBis}
\end{center}
\end{figure*}

\begin{figure*}
\centering
\includegraphics[width=0.3\linewidth]{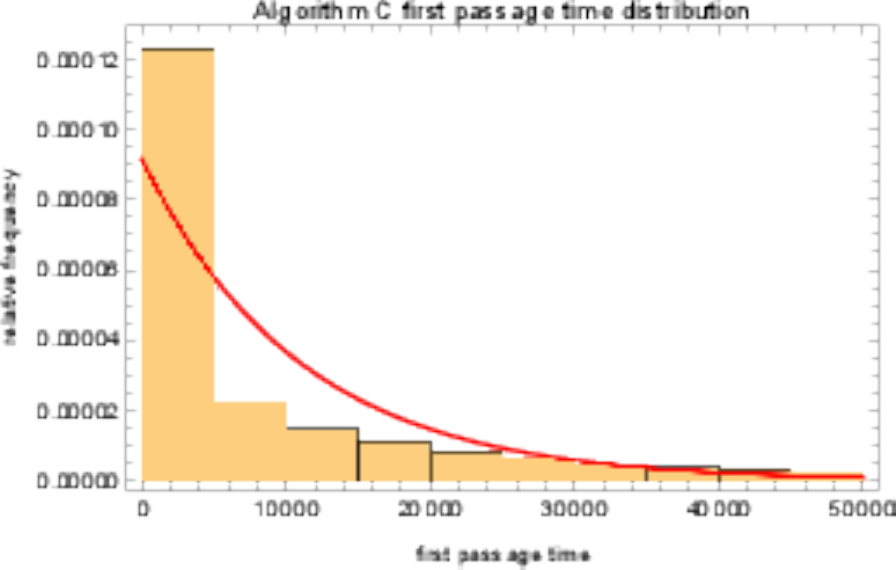}
\quad
\includegraphics[width=0.3\linewidth]{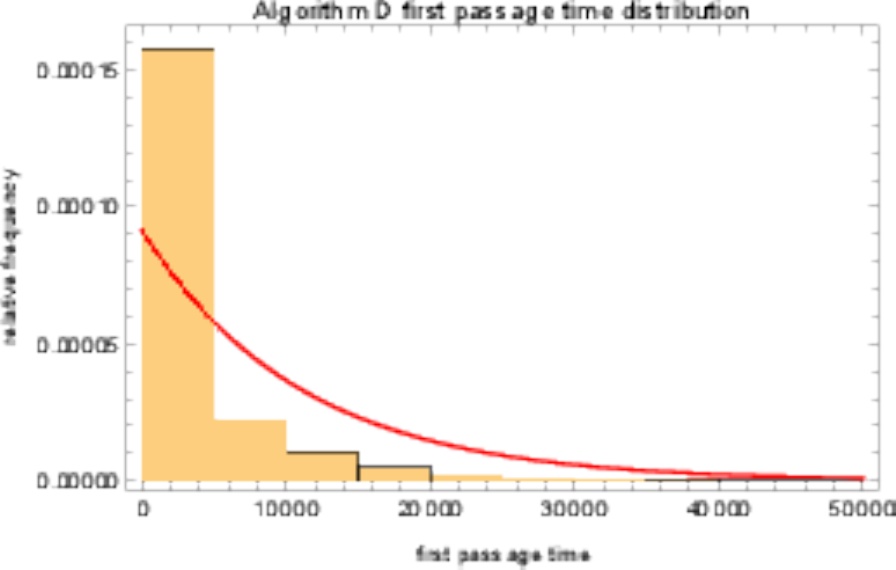}
\includegraphics[width=0.3\linewidth]{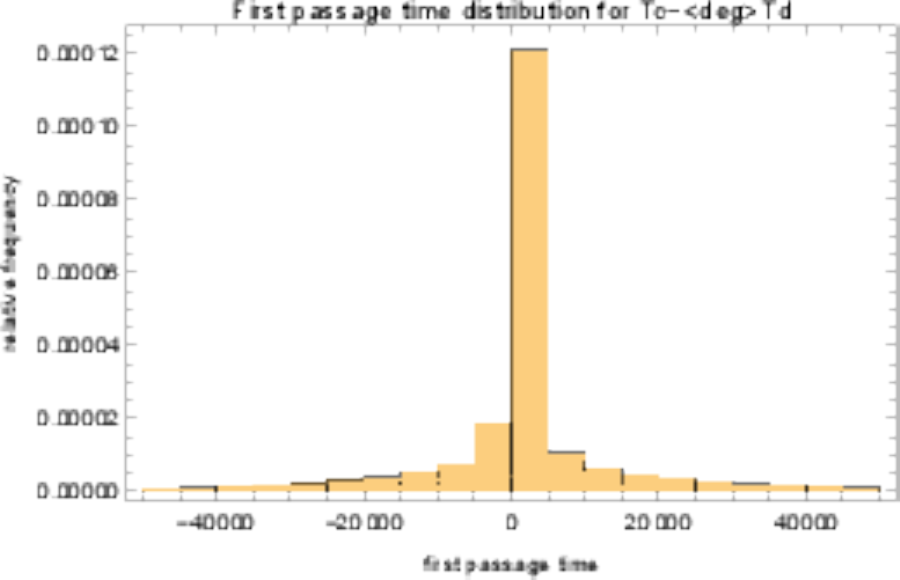}
\caption{Relative histograms for first passage times for the continuous-time,
event-driven Monte Carlo algorithm and the Gillespie algorithm.
(a) Probability distribution for Algorithm~3 trajectory time $T_c$ from the
empty initial structure to the MFE structure, with
mean of $10953.60$, standard deviation of $17765.47$,
maximum of $221,692.29$, and minimum of $0.41$. 
Red line shows the
exponential distribution with parameter $\lambda = \frac{1}{10953.60} \approx
9.13 \times 10^{-5}$.
(b) Probability distribution for Algorithm~4 trajectory time $T_d$ from the
empty initial structure to the MFE structure, with 
mean of $3192.31$, standard deviation of $5654.83$,
maximum of $72,473.93$, and minimum of $0.075$.
Red line shows the
exponential distribution with parameter $\lambda = \frac{1}{3192.31} \approx
3.13 \times 10^{-4}$ 
(c) Probability distribution for time $T_c - \langle N \rangle \cdot T_d$,
where $\langle N \rangle = 3.031162 \approx 3.03$ is the expected number
of neighbors for the network of all 62 secondary structures for the 10-nt 
sequence GGGGGCCCCC. This distribution has
mean of $1277.20$, standard deviation of $18,379.89$,
maximum of $204,747.05$, and minimum of $-199,785.20$.
See text for a proof that with overwhelming statistical significance we have
the strict inequalities
${\langle N \rangle}_b \cdot T_d < T_c < {\langle N \rangle}_u \cdot T_d$, 
where ${\langle N \rangle}_b$ [resp.  ${\langle N \rangle}_u$] denotes the
Boltzmann [resp. uniform] expected number of neighbors.
}
\label{fig:relativeHistogramAlgoCandDandCminusDegTimesD}
\end{figure*}

\hfill\break \clearpage
\begin{footnotesize}
\bibliographystyle{plain}

\end{footnotesize}

\hfill\break \clearpage
\noindent {\bf\large Appendix}
\appendix

\section{Markov chain of RNA secondary structures and detailed balance}

In this section, we prove that detailed balance does not hold for the
Markov chain $\mathcal{M}_1$ of secondary structures for an RNA sequence,
where transition probabilities are defined by 
equation~(\ref{eqn:transitionProb1}).
The failure of detailed balance is because for almost any RNA sequence,
there are secondary structures $x,y$ that are neighbors of each other,
but number $N(x)$ of neighbors of $x$ does not equal the
number $N(y)$ of neighbors of $y$.  If every secondary structure of
a given RNA sequence has the same number of neighbors, then it is easy
to show that the Boltzmann distribution $(\pi_1,\ldots,\pi_n)$  is
the stationary distribution, and a small calculation then shows that
detailed balance holds. If different structures have different numbers of
neighbors, then the Boltzmann distribution is {\em not} in general
the stationary distribution. To produce an example where detailed balance 
fails, we must compute the stationary distribution.

For convenience in following the
computation, we first present a simple example using 
the Nussinov energy model, in which the energy $E(s)$ of a structure is
defined to be $-1$ multiplied by the number of base pairs. Historically,
the Nussinov algorithm and  energy model \cite{nussinovJacobson} predate
the Zuker algorithm \cite{zukerStiegler} and  Turner
energy model \cite{Turner.nar10}.

\begin{example}[Detailed balance does not hold for Nussinov energy model]
Consider the 6-nt RNA sequence GGGCCC of length 6, which has only
four secondary structures, each having $N(s)$ neighbors, 
given in dot-bracket notation as follows:
\begin{enumerate}
\item
$\bullet \bullet \bullet \bullet \bullet \bullet$:  the empty structure $s_1$
with no base pairs and $N(s_1)=3$ neighbors.
\item
$\op \bullet \bullet \bullet \bullet \cp$: the structure $s_2$
with base pair $(1,6)$ and $N(s_2)=1$ neighbor.
\item
$\op \bullet \bullet \bullet \cp \bullet$: the structure $s_3$
with base pair $(1,5)$ and $N(s_3)=1$ neighbor.
\item
$\bullet \op \bullet \bullet \bullet \cp$: the structure $s_4$
with base pair $(2,6)$ and $N(s_4)=1$ neighbor.
\end{enumerate}
\end{example}
The transition probability matrix $M$ is given as follows, where we
compute the values to 20 decimal places using 
equation~(\ref{eqn:transitionProb1}), but display only 
4 decimal places for notational convenience:
\begin{eqnarray}
\label{eqn:transitionProbNussinovExample}
M &= \left( \begin{array}{llll}
0.0000&	0.333\overline{3} &0.333\overline{3} &0.333\overline{3}\\
0.1974& 0.8026& 0.0000 & 0.0000\\
0.1974& 0.8026& 0.0000 & 0.0000\\
0.1974& 0.8026& 0.0000 & 0.0000
\end{array} \right)
\end{eqnarray}
Let $M^{\infty} = \lim\limits_{n \rightarrow \infty} M^n$.
The stationary probability distribution for $M$ must satisfy
$M (p_1^*, p_2^*, p_3^*, p_4^*)^T = (p_1^*, p_2^*, p_3^*, p_4^*)^T$,
where $T$ denotes transpose, and so 
\begin{eqnarray*}
M^{\infty} &= \left( \begin{array}{llll}
p_1^*& p_2^*& p_3^* & p_4^*\\
p_1^*& p_2^*& p_3^* & p_4^*\\
p_1^*& p_2^*& p_3^* & p_4^*\\
p_1^*& p_2^*& p_3^* & p_4^*
\end{array} \right)
\end{eqnarray*}
Using the actual 20-place values of the transition probability matrix $M$,
whose 4-place approximations are given in
equation~(\ref{eqn:transitionProbNussinovExample}),
we obtain the millionth power of $M^{(10^6)}$ by a Mathematica computation.
The computed 16-place values are shown to 4-place accuracy in the following:
\begin{eqnarray*}
M^{(10^6)} &= \left( \begin{array}{llll}
0.1648& 0.2784& 0.2784& 0.2784\\
0.1648& 0.2784& 0.2784& 0.2784\\
0.1648& 0.2784& 0.2784& 0.2784\\
0.1648& 0.2784& 0.2784& 0.2784\\
\end{array} \right)
\end{eqnarray*}
so that the 4-place approximations of the stationary probabilities
are as follows:
$p_1^* = 0.1648$, 
$p_2^* = 0.2784$, 
$p_3^* = 0.2784$, 
$p_4^* = 0.2784$. 
We then compute that
$p_1^* \cdot p_{1,2} = 0.0550$, while 
$p_2^* \cdot p_{2,1} = 0.0617$, 
which shows that detailed balance does not hold.

Using the Turner energy model \cite{Turner.nar10}, 
the same example yields a different
transition probability matrix (not shown), whose millionth power is
as follows:
\begin{eqnarray*}
M^{(10^6)} &= \left( \begin{array}{llll}
0.9994655553038468& 0.000430085352242845&
0.0000521796443227893& 0.0000521796443227893\\
0.9994655553038468& 0.000430085352242845&
0.0000521796443227893& 0.0000521796443227893\\
0.9994655553038468& 0.000430085352242845&
0.0000521796443227893& 0.0000521796443227893\\
0.9994655553038468& 0.000430085352242845&
0.0000521796443227893& 0.0000521796443227893
\end{array} \right)
\end{eqnarray*}
We find that
$p_1^* =0.9994655553038468164 \approx 0.9995$,
$p_2^* =0.000430315331989701045645180244 \approx 0.0004$,
$p_{1,2}=0.001288878388504780024906293256 \approx 0.0013$,
$p_{2,1}=0.061737166286699882156163710079 \approx 0.617$,
and that
$p_1^* \cdot p_{1,2} = 0.0012881895542860573 \approx 0.00129$ while
$p_2^* \cdot p_{2,1} = 0.00002656644920676464 \approx 0.00003$.
It follows that detailed balance does not hold for the Turner energy
model. For RNA sequences of length $n$, there are exponentially many
($\approx 1.95947  \cdot n^{-3/2} \cdot 1.86603^n$) secondary structures,
so similar computations cannot be carried out, and it may be the case that
detailed balance ``approximately'' holds.

\end{document}